\documentstyle[12pt]{article}
\let\rf=\rfloor
\def\rfloor{\rf_{\! q}}
\def\gtil{{\tilde g}}
\def\demi{{{1\over 2}}}
\def\xii{\xi^{({1\over 2})}}

\def\psii{\psi^{({1\over 2})}}

\def\Jpo{{\cal O}\! \left[J_+\right]}
\def\Jmo{{\cal O}\! \left[J_-\right]}
\def\Jpmo{{\cal O}\! \left[J_\pm \right]}
\def\Jto{{\cal O}\! \! \left[q^{J_3}\right]}

\def\Tpo{{\cal O}\! \left[T^+\right]}
\def\Tmo{{\cal O}\! \left[T^-\right]}
\def\Tpmo{{\cal O}\! \left[T^\pm \right]}
\def\Tto{{\cal O}\! \! \left[q^{T_3}\right]}
\def\Tso{{\cal O}\! \! \left[q^{-T_3}\right]}
\def\Ttso{{\cal O}\! \! \left[q^{\pm T_3}\right]}
\def\Tsto{{\cal O}\! \! \left[q^{\mp T_3}\right]}
\def\Do{{\cal O}\! \! \left[D\right]}

\def\Jpop#1{{\cal O}\! \left[J_+\right]_{#1}^{(R)}}
\def\Jmop#1{{\cal O}\! \left[J_-\right]_{#1}^{(R)}}
\def\Jtop#1{{\cal O}\! \! \left[q^{J_3}\right]_{\! #1}^{(R)}}

\def\Dop#1,#2{{\cal O}\! \left[D\right]_{#1,#2}^{(R)}}

\def\Jtpsi#1,#2,#3,#4,#5{[ q^{{\cal J}_3, #2}_{\phantom{{\cal J}_3}
#5}{}]^{(#1)}_{#3#4}}
\def\Jspsi#1,#2,#3,#4,#5{[ q^{-{\cal J}_3, #2}_{\phantom{-{\cal J}_3}
#5}{}]^{(#1)}_{#3#4}}
\def\Jtspsi#1,#2,#3,#4,#5{[ q^{\pm{\cal J}_3, #2}_{\phantom{-{\cal
J}_3}
#5}{}]^{(#1)}_{#3#4}}
\def\Jppsi#1,#2,#3,#4,#5{[ {\cal J}^{\phantom{+,}#2}_{+,
#5}{}]^{(#1)}_{#3#4}}
\def\Jpmpsi#1,#2,#3,#4,#5{[ {\cal J}^{\phantom{\pm,}#2}_{-,
#5}{}]^{(#1)}_{#3#4}}
\def\Jts#1,#2,#3,#4,#5{[ q^{\pm{ J}_3}
{}]_{#3#4}}
\def\Jpm#1,#2,#3,#4,#5{[  J_\pm ^{\phantom{\pm,}}
{}]_{#3#4}}
\def\UU#1,#2,#3,#4,#5{  {\cal U}
^{(#1)}_{#3#4}\! (#2)}
\def\VV#1,#2,#3,#4,#5{  {\cal V}
^{(#1)}_{#3#4}\! (#2)}

\def\U#1,#2,#3,#4{{\cal U}^{(#1)}_{#2 #3} (#4)}
\def\V#1,#2,#3,#4{{\cal V}^{(#1)}_{#2 #3} (#4)}

\def\journal#1, #2, #3, #4 { {\sl #1~}{\bf #2~} (#3)  #4 }

\def\cmp{\journal Comm. Math. Phys., }

\def\np{\journal Nucl. Phys., }

\def\pl{\journal Phys. Lett., }

\def\ijmp{\journal Int. J. Mod. Phys., }

\catcode`\@=11
\def\marginnote#1{}
\newcount\hour
\newcount\minute
\newtoks\amorpm
\hour=\time\divide\hour by60
\minute=\time{\multiply\hour by60 \global\advance\minute
by-\hour}\edef\standardtime{{\ifnum\hour<12
\global\amorpm={am}%
        \else\global\amorpm={pm}\advance\hour by-12 \fi
        \ifnum\hour=0 \hour=12 \fi
        \number\hour:\ifnum\minute<10
0\fi\number\minute\the\amorpm}}
\edef\militarytime{\number\hour:\ifnum\minute<10
0\fi\number\minute}

\def\draftlabel#1{{\@bsphack\if@filesw {\let\thepage\relax
   \xdef\@gtempa{\write\@auxout{\string
      \newlabel{#1}{{\@currentlabel}{\thepage}}}}}\@gtempa
   \if@nobreak \ifvmode\nobreak\fi\fi\fi\@esphack}
        \gdef\@eqnlabel{#1}}
\def\@eqnlabel{}
\def\@vacuum{}
\def\draftmarginnote#1{\marginpar{\raggedright\scriptsize\tt#1}}
\def\draft{\oddsidemargin -.5truein
        \def\@oddfoot{\sl preliminary draft \hfil
        \rm\thepage\hfil\sl\today\quad\militarytime}
        \let\@evenfoot\@oddfoot \overfullrule 3pt
        \let\label=\draftlabel
        \let\marginnote=\draftmarginnote

\def\@eqnnum{(\theequation)\rlap{\kern\marginparsep\tt\@eqnlabel}%
\global\let\@eqnlabel\@vacuum}  }

\def\numberbysection{\@addtoreset{equation}{section}
        \def\theequation{\thesection.\arabic{equation}}}

\def\underline#1{\relax\ifmmode\@@underline#1\else
 $\@@underline{\hbox{#1}}$\relax\fi}

\catcode`@=12
\relax
\numberbysection
\def\beq{\begin{equation}}
\def\eeq{\end{equation}}
\def\beqa{\begin{eqnarray}}
\def\eeqa{\end{eqnarray}}
 \def\nnn{\nonumber \\}

\def\hhat{{\widehat h}}
\def\qhat{{widehat q}}

\def\Jhat{{\widehat J}}

\def\qhat{{\widehat q}}

\def\Jgen#1 {  {\underline J_{#1}} }
\def\Jgenp#1 #2 {(J_{#1}+{#2},\Jhat_{#1})}
\def\Jgenm#1 #2 {(J_{#1}-{#2},\Jhat_{#1})}
\def\Jg#1 {J_{#1},\Jhat_{#1}}
\def\Jgp#1 #2 {J_{#1}+{#2},\Jhat_{#1}}
\def\Mgen#1 {{\underline M_{#1}}}
\def\mgen#1 {{\underline m_{#1}}}

\def\fin{\end{document}}

\def\Jgen#1 {  {\underline J_{#1}} }
\def\Jgenp#1 #2 {(J_{#1}+{#2},\Jhat_{#1})}
\def\Jgenm#1 #2 {(J_{#1}-{#2},\Jhat_{#1})}
\def\Jg#1 {J_{#1},\Jhat_{#1}}
\def\Jgp#1 #2 {J_{#1}+{#2},\Jhat_{#1}}
\def\Mgen#1 {{\underline M_{#1}}}

\def\fusV#1,#2,#3,#4,#5,#6 {f_V(
\Jgen{#1} ,
\Jgen{#2} ,
\Jgen{#3} ,
\Jgen{#4} ,
\Jgen{#5} ,
\Jgen{#6} )}

\def\demi{{{1\over 2}}}
\def\xii{\xi^{({1\over 2})}}

\def\psii{\psi^{({1\over 2})}}

\def\Jpo{{\cal O}\! \left[J_+\right]}
\def\Jmo{{\cal O}\! \left[J_-\right]}
\def\Jpmo{{\cal O}\! \left[J_\pm \right]}
\def\Jto{{\cal O}\! \! \left[q^{J_3}\right]}

\def\Jtpsi#1,#2,#3,#4,#5{[ q^{{\cal J}_3, #2}_{\phantom{{\cal J}_3}
#5}{}]^{(#1)}_{#3#4}}
\def\Jspsi#1,#2,#3,#4,#5{[ q^{-{\cal J}_3, #2}_{\phantom{-{\cal J}_3}
#5}{}]^{(#1)}_{#3#4}}
\def\Jppsi#1,#2,#3,#4,#5{[ {\cal J}^{\phantom{+,}#2}_{+,
#5}{}]^{(#1)}_{#3#4}}
\def\Jmpsi#1,#2,#3,#4,#5{[ {\cal J}^{\phantom{-,}#2}_{-,
#5}{}]^{(#1)}_{#3#4}}

\def\U#1,#2,#3,#4{{\cal U}^{(#1)}_{#2 #3} (#4)}
\def\V#1,#2,#3,#4{{\cal V}^{(#1)}_{#2 #3} (#4)}

\def\demi{{{1\over 2}}}
\def\xii{\xi^{({1\over 2})}}
\def\xic{\xi^{[\demi,\demi](0)}_0}

\def\psii{\psi^{({1\over 2})}}

\def\Jpo{{\cal O}\! \left[J_+\right]}
\def\Jmo{{\cal O}\! \left[J_-\right]}
\def\Jpmo{{\cal O}\! \left[J_\pm \right]}
\def\Jto{{\cal O}\! \! \left[q^{J_3}\right]}

\def\Tpo{{\cal O}\! \left[T_+\right]}
\def\Tmo{{\cal O}\! \left[T_-\right]}
\def\Tpmo{{\cal O}\! \left[T_\pm \right]}
\def\Tto{{\cal O}\! \! \left[q^{T_3}\right]}
\def\Tso{{\cal O}\! \! \left[q^{-T_3}\right]}
\def\Ttso{{\cal O}\! \! \left[q^{\pm T_3}\right]}
\def\Tsto{{\cal O}\! \! \left[q^{\mp T_3}\right]}
\def\Do{{\cal O}\! \! \left[D\right]}

\def\Jpop#1{{\cal O}\! \left[J_+\right]_{#1}}
\def\Jmop#1{{\cal O}\! \left[J_-\right]_{#1}}
\def\Jpmop#1{{\cal O}\! \left[J_\pm \right]_{#1}}
\def\Jtop#1{{\cal O}\! \! \left[q^{J_3}\right]_{\! #1}}

\def\Dop#1,#2{{\cal O}\! \left[D\right]_{#1,#2}}

\def\Top#1,#2,#3{{\cal O}\! \left[T_{#1 2 #2}\right]_{#3}}
\def\Ttop#1{{\cal O}\! \left[q^{T_3}\right]}
\def\Tsop#1{{\cal O}\! \left[q^{-T_3}\right]}
\def\Ttsop#1{{\cal O}\! \left[q^{\pm T_3}\right]}
\def\Tpop#1{{\cal O}\! \left[T_+\right]_{#1}}
\def\Tmop#1{{\cal O}\! \left[T_-\right]_{#1}}
\def\Tpmop#1{{\cal O}\! \left[T_\pm\right]_{#1}}

\def\Jtpsi#1,#2,#3,#4,#5{[ q^{{\cal J}_3, #2}_{\phantom{{\cal
J}_3}
#5}{}]^{(#1)}_{#3#4}}
\def\Jttpsi#1,#2,#3,#4,#5{[ q^{{2\cal J}_3, #2}_{\phantom{{\cal
J}_3}
#5}{}]^{(#1)}_{#3#4}}
\def\Jspsi#1,#2,#3,#4,#5{[ q^{-{\cal J}_3, #2}_{\phantom{-{\cal
J}_3}
#5}{}]^{(#1)}_{#3#4}}
\def\Jsspsi#1,#2,#3,#4,#5{[ q^{-{2
\cal J}_3, #2}_{\phantom{-{\cal J}_3}
#5}{}]^{(#1)}_{#3#4}}
\def\Jppsi#1,#2,#3,#4,#5{[ {\cal J}^{\phantom{+,}#2}_{+,
#5}{}]^{(#1)}_{#3#4}}
\def\Jpmpsi#1,#2,#3,#4,#5{[ {\cal J}^{\phantom{+,}#2}_{\pm ,
#5}{}]^{(#1)}_{#3#4}}
\def\Xppsi#1,#2,#3,#4,#5{[ {\cal X}^{\phantom{+,}#2}_{+,
#5}{}]^{(#1)}_{#3#4}}
\def\Xpmpsi#1,#2,#3,#4,#5{[ {\cal X}^{\phantom{+,}#2}_{\pm ,
#5}{}]^{(#1)}_{#3#4}}
\def\Jmpsi#1,#2,#3,#4,#5{[ {\cal J}^{\phantom{-,}#2}_{-,
#5}{}]^{(#1)}_{#3#4}}
\def\Xmpsi#1,#2,#3,#4,#5{[ {\cal X}^{\phantom{-,}#2}_{-,
#5}{}]^{(#1)}_{#3#4}}
\def\Tpsi#1,#2,#3,#4,#5{[ {\cal T}^{#2}_{
#5}{}]^{(#1)}_{#3,#4}}
\def\A#1,#2,#3,#4,#5{[ { A}^{#2}_{
#5}{}]^{(#1)}_{#3,#4}}
\def\B#1,#2,#3,#4,#5{[ { B}^{#2}_{
#5}{}]^{(#1)}_{#3,#4}}
\def\An#1,#2,#3,#4,#5,#6{[ {A}^{#2}]^{(#1)}
_{{#3,#4} ; {#5,#6}}}
\def\Bn#1,#2,#3,#4,#5,#6{[ {B}^{#2}]^{(#1)}
_{{#3,#4} ; {#5,#6}}}
\def\CA#1,#2,#3,#4,#5{[ q^{{\cal A}_{3}}]^{(#1)}_{
#2,#3;#4,#5}}
\def\CB#1,#2,#3,#4,#5{[ q^{{\cal B}_{3}}]^{(#1)}_{
#2,#3;#4,#5}}
\def\CAn#1,#2,#3,#4,#5,#6{[ {\cal A}_{#2}]^{(#1)}
_{{#3,#4} ; {#5,#6}}}
\def\CBn#1,#2,#3,#4,#5,#6{[ {\cal B}_{#2}]^{(#1)}
_{{#3,#4} ; {#5,#6}}}

\def\LJtpsi#1,#2,#3,#4,#5,#6,#7,#8{
[ \Lambda\left(q^{{\cal J}_3, #3}\right)_{
#8}{}]^{(#1,#2)}_{#4#5\,#6#7}}
\def\LJspsi#1,#2,#3,#4,#5,#6,#7,#8{
[ \Lambda\left(q^{-{\cal J}_3, #3}\right)_{
#8}{}]^{(#1,#2)}_{#4#5\,#6#7}}
\def\LJppsi#1,#2,#3,#4,#5,#6,#7,#8{
[ \Lambda\left({\cal J_+}\right)^{#3}_{
#8}{}]^{(#1,#2)}_{#4#5\,#6#7}}
\def\LJmpsi#1,#2,#3,#4,#5,#6,#7,#8{
[ \Lambda\left({\cal J_-}\right)^{#3}_{
#8}{}]^{(#1,#2)}_{#4#5\,#6#7}}
\def\LJpmpsi#1,#2,#3,#4,#5,#6,#7,#8{
[ \Lambda\left({\cal J_\pm}\right)^{#3}_{
#8}{}]^{(#1,#2)}_{#4#5\,#6#7}}
\def\LXppsi#1,#2,#3,#4,#5,#6,#7,#8{
[ \Lambda\left({\cal  X_+}\right)^{#3}_{
#8}{}]^{(#1,#2)}_{#4#5\,#6#7}}
\def\LXmpsi#1,#2,#3,#4,#5,#6,#7,#8{
[ \Lambda\left({\cal X_-}\right)^{#3}_{
#8}{}]^{(#1,#2)}_{#4#5\,#6#7}}
\def\LXpmpsi#1,#2,#3,#4,#5,#6,#7,#8{
[ \Lambda\left({\cal  X_\pm}\right)^{#3}_{
#8}{}]^{(#1,#2)}_{#4#5\,#6#7}}
\def\LTpsi#1,#2,#3,#4,#5,#6,#7,#8{
[ \Lambda\left({\cal T}\right)^{#3}_{
#8}{}]^{(#1,#2)}_{#4#5\,#6#7}}

\def\U#1,#2,#3,#4{{\cal U}^{(#1)}_{#2 #3} (#4)}
\def\V#1,#2,#3,#4{{\cal V}^{(#1)}_{#2 #3} (#4)}

\def\pish{{\pi\over h}}
\begin{document}

\begin{titlepage}

\nopagebreak \begin{flushright}

LPTENS--96/20\\
CERNTH/96-94\\
hep-th/9604131
 \\
    April    1996
\end{flushright}

\vglue 2  true cm
\begin{center}
{\large \bf
HIDDEN  U$_{\hbox{\bf  q}}$ (sl(2))
${\bf \otimes}$ U$_{\hbox{\bf q}}$ (sl(2))    \\
\medskip
QUANTUM GROUP SYMMETRY\\
\medskip
IN TWO DIMENSIONAL GRAVITY\footnote{Work supported in part by the
Human Capital
and Mobility Network EC programme ERBCHRCT920035, and the
S.C.I.E.N.C.E. EC programme
SC1*CT920789.}}
\vglue 1.5 true cm
{\bf Eug\`ene CREMMER}\\
\medskip
{\bf Jean-Loup~GERVAIS}\\
\medskip
{\footnotesize Laboratoire de Physique Th\'eorique de
l'\'Ecole Normale Sup\'erieure\footnote{Unit\'e Propre du
Centre National de la Recherche Scientifique,
associ\'ee \`a l'\'Ecole Normale Sup\'erieure et \`a
l'Universit\'e
de Paris-Sud.},\\
24 rue Lhomond, 75231 Paris CEDEX 05, ~France},\\
{\bf Jens SCHNITTGER}\\
{\footnotesize Theory Division, CERN.}\\
\medskip
\end{center}
\vglue 2 true cm
\begin{abstract}
\baselineskip .4 true cm
{\footnotesize
\noindent
In a previous paper, the quantum-group-covariant chiral vertex
operators in the
spin $1/2$ representation were shown to act, by braiding with the
other covariant primaries,
as   generators of the well known $U_q(sl(2))$ quantum group
symmetry (for a single screening charge).  Here, this structure is
transformed  to the
Bloch wave/Coulomb gas operator
basis, thereby establishing for the first time its quantum group
symmetry properties.
 A  $U_q(sl(2))\otimes U_q(sl(2))$ symmetry of a
novel type emerges: The two Cartan-generator eigenvalues
are specified by the choice of  matrix element (bra/ket
Verma-modules); the two
Casimir eigenvalues
are equal and specified by the Virasoro weight of the vertex operator
considered; the co-product is defined with a matching condition
dictated by the
Hilbert space structure of the  operator product.  This hidden
symmetry
possesses  a novel
Hopf like structure
compatible with  these  conditions.
At roots of unity it gives the right truncation. Its (non linear)
connection with the
 $U_q(sl(2))$ previously discussed is disentangled.
}
\end{abstract}
\vfill
\end{titlepage}
\section{Introduction}
Quantum integrability as we know it is essentially synonymous to the
concept
of  $R$ matrix and   Yang-Baxter relations. While it is not known
whether
the latter always possess a group-theoretical interpretation, it is
widely
believed that this is true at least for the subclass of conformal
integrable
systems; well-known examples are given by the minimal models,
the WZW models and Liouville/Toda theory, where
the underlying symmetries are indeed known to be given by
quantum groups \cite{FFK}\cite{AGS}\cite{B}.
However, in spite of extensive studies \cite{MR}\cite{AFT}\cite{HM},
our understanding of the
quantum group symmetry in these theories is still
somewhat incomplete as we lack an explicit realization of the
symmetry generators
as operators on the Hilbert space, similar to the representation of
classical
symmetry groups in conventional field theory. In a first paper
\cite{CGS1}, we have analyzed
this question within the context of 2d gravity,
and proposed a novel approach
involving position-dependent symmetry generators. Perhaps the most
striking feature of
these generators
is the fact that they are given in terms of the same operators that
form irreducible
representations of the quantum group $U_q(sl(2))$; more precisely,
the basic generators $J_\pm, J_3$ were seen to be related to the
fundamental representation
of spin $\demi$, while the full enveloping algebra arises from tensor
products of the
latter and thus involves all the higher spins.

The covariant operator basis
 \cite{B} used in ref. \cite{CGS1}  has the appealing property of
consisting
of  conformal primaries only, while in previous work it was found
necessary to introduce
vertex operators that are not fully covariant under the Virasoro
symmetry \cite{GS}.
The latter, on the other hand, have a somewhat simpler structure as
they are
given directly in terms of the familiar Coulomb gas vertex operators
used for the free
field description  of rational conformal field theories.
Remarkably, a conformally covariant version of the latter is
 known to constitute an alternative, equivalent operator basis for
the description of 2d
gravity, which we will call the Bloch wave basis\footnote{This name
is motivated
by the fact that these fields have well-defined monodromy
properties.}; however,
 "conformal covariantization" does not conserve the transformation
behaviour of these vertex operators under the quantum group. The
question therefore
arises naturally how the quantum
group symmetry is realized in terms of the Bloch wave operators, and
what is
the relation between the symmetry generators in both pictures; this
is the basic
theme of the present paper. {}From a Hamiltonian point of view, one
would  expect
 that the symmetry generators are simply the same, as both sets
of fields are living in the same Hilbert space. However,  it turns
out that
there is a more natural realization of the quantum group symmetry on
the Bloch waves, where
the generators are also given by Bloch wave fields. Here
  the free field zero mode will play
the role of the generator $J_3$ in the Bloch wave basis, just as in
the conventional
description,
while $J_\pm$  will be given in terms
of the Bloch wave operators of spin $\demi$.

The structure we find in the Bloch wave basis proves to be much
more  intricate than in the covariant one: On the one hand, the
commutation relations of
the generators turn
out to be essentially (up to central terms)
 those of $U_{\sqrt q}(sl(2))$, rather than $U_q(sl(2))$. On the
other
hand, these generators induce a   symmetry of the operator
algebra of the Bloch waves which is larger than the one of their
commutation relations.
This   ``internal'' symmetry  has
a natural description in terms of  $U_q(sl(2))\otimes U_q(sl(2))$
\footnote{
Note that this structure has nothing to do with the $U_q(sl(2))\odot
U_{\qhat}(sl(2))$ discussed in ref. \cite{CGS1}, which arises from
considering both semiclassical and non-semiclassical deformations
of $sl(2)$, or in the language of $2d$ gravity, both screening
charges. In the
present paper, we will restrict ourselves to a single screening
charge throughout.}.
We carefully show, using our explicit constructions,  how these three
algebras are
related, both $U_q(sl(2))$ and  $U_{\sqrt q}(sl(2))$ being non
linearly
``embedded'' in our $U_q(sl(2))\otimes U_q(sl(2))$, which is of a
novel type. The
fact that the Bloch waves are intertwining operators from the
conformal point of view leads
to two additional constraints: There is a matching condition
for the magnetic indices of the two factors, and the Casimir
eigenvalues
of the two representations involved must be equal. Remarkably, there
exists
a new Hopf like  structure, different from the standard one of
$U_q(sl(2))\otimes
U_q(sl(2))$, which is consistent with these conditions.

The paper is organized as follows. In section \ref{covbasis}, we
recall the essential
points of the
analysis of the first paper, and anticipate the general form of the
relation between
the generators in the two pictures. In section 3, we work out
explicitly
the passage from the covariant to the Bloch wave generators by means
of a nonlinear
redefinition, and establish the commutation relations of the latter.
 In section 4, we show that the action of the new generators on the
Bloch waves can
be described in terms of two commuting sets of matrices, both of
which fulfill
the commutation relations of $U_q(sl(2))$. We work out the new
coproduct
structure induced by the intertwining constraints, which makes this
matrix
algebra different from $U_q(sl(2))\otimes U_q(sl(2))$ as a Hopf
algebra.
We characterize the symmetries of the operator algebra generated by
these
matrices and explain that the q $6j$ symbols
describing the fusion of the Bloch wave fields can be viewed as
Clebsch-Gordan
 coefficients for the new symmetry structure.
In section 5, we will show how this new symmetry directly
connected with the Liouville zero mode can be used to
classify the spectrum of primary fields and associated
Verma modules. We will also discuss partially the case of
$q$ root of unity and show that the corresponding
representation of $U_q(sl(2))\otimes U_q(sl(2))$
gives the right truncation for the spectrum
of zero modes.
 As an application of the formalism,
we derive  the transformation laws of the Coulomb gas
operators heavily used in previous work on the quantum Liouville
theory.
In section 6  we  depart from the Liouville system and consider a
conformal
theory, as yet unknown, where the full  $U_q(sl(2))\otimes
U_q(sl(2))$ would be
operatorially realized. Some reasonable hypothesis allow us to derive
the algebra of the
quantum generators, which is a central extension of this internal
symmetry group.
Finally, in section 7 we derive the full Hopf like structure using
the properties
of the operator algebra as a guide. Although it does not
strictly obey the usual axioms,  a coproduct, a counit and an
antipode may be defined
which satisfy very natural counterparts of the standard relations.
 We close with some  open questions and indications of possible
further
developments along the lines of the present analysis.


\section{Quantum Group Action in the Covariant Basis}
\label{covbasis}
We begin with a short recapitulation of the main results of ref.
\cite{CGS1}.
The starting point is the operator algebra of the chiral primaries
$\xi_M^{(J)}$,
which were constructed in refs.\cite{B} and shown to form
representations of spin $J$
of $U_q(sl(2))$. The basic observation now is that the special
operators
$\xi^{(\demi)}_{\pm\demi}$ can be viewed not only as covariant
fields, but at the same time as generators. We define
  the following set of operators:
\beq
\Jpo_{\sigma_+} \equiv\kappa^{(+)}_+ \xi_{{1\over 2}}^{({1\over
2})}(\sigma_+), \quad
\Jto_{\sigma_+} \equiv \kappa^{(+)}_3 \xi_{-{1\over 2}}
^{({1\over
2})}(\sigma_+),
\label{Jopdef+}
\eeq
where $\kappa^{(+)}_+$ and $\kappa^{(+)}_3$  are suitable
normalization
 constants\footnote{In the previous paper, they have been denoted
$\kappa_+^{(R+)}, \kappa_3^{(R+)}$. Since we will consider only the
action
of the generators to the right in this paper, the index $R$ is
dropped
here and below.}, related by $
\kappa^{(+)}_+ / \kappa^{(+)}_3  =
 q^{\demi}/ (1-q^2)$
and similarly
\beq
\Jmo_{\sigma_-} \equiv\kappa^{(-)}_- \xi_{-{1\over 2}}^{({1\over
2})}(\sigma_-), \quad
\Jto_{\sigma_-} \equiv \kappa^{(-)}_3 \xi_{ {1\over 2}}^{({1\over
2})}(\sigma_-),\quad
\label{Jopdefr-}
\eeq
{}From the braiding matrix of the $\xi^{(J)}_M$
fields, namely the universal $R$ -matrix of $U_q(sl(2))$, it follows
immediately  that
\beq
\Jto_{\sigma_\pm} \xi_{M}^{(J)}(\sigma)=
  \xi_{N}^{(J)}(\sigma)  \left[ q^{J_3}\right ]_{NM}
\Jto_{\sigma_\pm}
\label{act3+}
\eeq
\beq
\Jpmo_{\sigma_\pm} \xi_{M}^{(J)}(\sigma)=
 \xi_{N}^{(J)}(\sigma) \left[q^{-J_3}\right]_{NM}\Jpmo_{\sigma_\pm} +
\xi_{N}^{(J)}(\sigma) \left[J_\pm\right]_{NM} \Jto_{\sigma_\pm},
\label{act+}
\eeq
whenever $\sigma_+> \sigma$,
resp. $\sigma_-<\sigma$.
Here $\left[ J_\pm\right]_{NM}$ and $\left[ q^{J_3}\right]_{NM}$
denote
the usual representation matrices of the $U_q(sl(2))$ generators.
Eqs. \ref{act3+}--\ref{act+} describe  the action
of $U_q(sl(2))$ by coproduct, in accord with the general framework
exposed in
ref. \cite{MaSc}. Indeed, the $U_q(sl(2))$ coproduct
$
\Lambda(q^{\pm J_3})=q^{\pm J_3}\otimes q^{\pm J_3}$,
$\Lambda(J_\pm)=J_\pm \otimes q^{J_3}+q^{-J_3}\otimes J_\pm$,
is immediately recognized to appear in Eqs. \ref{act3+}, \ref{act+},
one
of the generators being realized as a matrix, and the other one as an
operator
in the Hilbert space.
Similarly, the action of the generators on products of fields is
given by
repeated application of the coproduct. On the other hand, the
commutation relations
of these generators were found to differ somewhat from the standard
$U_q(sl(2))$ ones,
essentially by central charges. They take the general form
\beq
 q\Jpo \Jto -\Jto \Jpo  = C_+
\label{defcp1}
\eeq
\beq
\Jto \Jmo  -q^{-1} \Jmo \Jto = C_-
\label{defcm1}
\eeq
\beq
 \Jpo \Jmo -\Jmo \Jpo =\Do +
 {(\Jto )^2 \over q- q^{-1}},
\label{defbeta1}
\eeq
where $\Do$ satisfies
\beq
C_+\Jmo
+C_-\Jpo  =\Do  \Jto
\label{D2}
\eeq
and $C_+, C_-$ are central charges. It was shown that this implies
in particular that we can reexpress $\Jpmo$ in terms of $\Do$
and $\Jto$ and that there exist some relations between
$\Do$ and $\Jto$
\beq
\Jto \Do -q^{\mp 1} \Do \Jto =\pm C_\mp (q-q^{-1}) \Jpmo.
\label{Dcom}
\eeq
\beq
\Jto \Do ^2 -(q+q^{-1}) \Do \Jto \Do +\Do ^2 \Jto  =
C_+C_-\Jto ,
\label{consist1}
\eeq
\beq
\Do \Jto ^2 -(q+q^{-1}) \Jto  \Do \Jto +\Jto ^2 \Do
=-C_+C_-(q-q^{-1}),
\label{consist2}
\eeq
While on a purely
formal level Eqs. \ref{defcp1} -- \ref{defbeta1} can be transformed
into the standard
$U_q(sl(2))$ commutation relations by redefinitions of the generators
that preserve the coproduct
action, this turns out to be impossible in our field-theoretic
realization where
the central charges are actually nontrivial operators. Thus, the
commutation
relations are realized only in this weak sense. Another peculiarity
of the
field-theoretic realization is that ordinary commutators are to be
replaced by
what we call fixed point (FP) commutators. Indeed, the operatorial
realization
of Eq. \ref{defcp1} is given by
\beq
 \Jto_{\sigma_+} \Jpo_{\sigma_+'} -q
\Jpo_{\sigma_+} \Jto_{\sigma'_+} =
-q^\demi \kappa^{(+)}_+\kappa^{(+)}_3 \>
\xi^{[\demi , \demi](0)}_0(\sigma_+, \sigma'_+),
\label{central+}
\eeq
where the central charge
$
\xi^{[\demi , \demi](0)}_0(\sigma_+, \sigma'_+)=
\sum_{M}
(\demi,M;\demi,-M|0)\xi_M^{(\demi)}(\sigma_+)\xi_{-M}^{(\demi)}
(\sigma'_+)$
commutes with all the $\xi$ fields being
the singlet formed from the product of two spin $\demi$
representations. Note that in Eq. \ref{central+},
 the arguments $\sigma_+,\sigma'_+$
of the operators are not exchanged, and this is precisely the meaning
of the FP prescription. The number of positions that appears in a
given
product of generators can be thought of as some kind of additive
gradation
of the formal algebra eqs. \ref{defcp1} -- \ref{defbeta1},
such that Eqs. \ref{defcp1}, \ref{defcm1}, \ref{defbeta1}
have grading $2$ and Eq. \ref{D2} grading $3$. \footnote{
Actually the relations of grading larger than two are not directly
FP realized in our scheme, but in the special case $C_+=-C_- ,
\Do \sim C_+$ we have here, cancellations occur which always allow
to reduce the grading to less than or equal to $2$.}
For $\sigma< \sigma_+, \sigma'_+$, it follows directly from the
braiding relations, governed by the universal $R$ matrix, that
the quantity $ \xi^{[\demi , \demi](0)}_0
(\sigma_+, \sigma'_+)$ commutes
with all $\xi^{(J)}_M(\sigma)$ and thus,
comparing with Eq.\ref{act3+}, one identifies
\beq
C_+= q^\demi \kappa^{(+)}_+\kappa^{(+)}_3 \>
\xi^{[\demi , \demi](0)}_0(\sigma_+, \sigma_+').
\label{Cplus}
\eeq
The discussion for the other Borel subalgebra
takes the same form with the obvious replacements, and one has
$C_-=-C_+$.
In order to write the FP commutator of $\Jpo$ and $\Jmo$, we have to
define
$\Jmo$ at points $\sigma_+,\sigma_+'$ (or $\Jpo$ at points $\sigma_-,
\sigma'_-$). This is achieved by invoking the monodromy operation
$\sigma \to \sigma +2\pi$, which transforms the point $\sigma_- <
\sigma$
into a point $\sigma_+ >\sigma$. The consistent definition is
$
\Jmo_{\sigma_+}=\kappa_-^{(+)}\xii_{-\demi}(\sigma_+-2\pi),
$
which  gives,  using the monodromy,
\beq
\Jmo_{\sigma_+}=\kappa_-^{(+)} \left [
q^{-\demi}(q^\varpi +q^{-\varpi})\xi^{(\demi)}_{-\demi}(\sigma_+)
-q^{-1}\xi^{(\demi)}_\demi(\sigma_+) \right]
\label{monodr}
\eeq
with
$
\kappa_-^{(+)} / \kappa^{(+)}_3 =
 q^{-\demi} /(1-q^{-2})$.
 $\varpi$ is the zero mode of the free field underlying the
construction
of the $\xi^{(J)}_M$ fields.
We remark that the zero mode $\varpi$ does not enter the algebra of
the
$\xi$ fields, which is given exclusively in terms of quantum group
symbols,
and thus the monodromy is the only place where it appears. Using
Eq.\ref{monodr}
one then obtains a realization of Eqs. \ref{D2}, \ref{defbeta1}:
\beq
C_+(\sigma_+',\sigma_+'')\>(\Jmo_{\sigma_+}
-\Jpo_{\sigma_+}) =
\Do_{\sigma_+',\sigma_+''}
\Jto_{\sigma_+}
\label{D2FP}
\eeq
and
\beq
 \Jpo_{\sigma_+}  \Jmo_{\sigma'_+} -\Jmo_{\sigma_+}
\Jpo_{\sigma'_+} =
\Do_{\sigma_+,\sigma_+'} +
 {\Jto_{\sigma_+} \Jto_{\sigma'_+}  \over q- q^{-1}} ,
\label{J+J-sigma}
\eeq
with
\beq
\Do_{\sigma_+,\sigma_+'} ={1\over  q-q^{-1}} 2\cos
(h\varpi) C_+(\sigma_+, \sigma_+'),
\label{Ddef}
\eeq
This shows that the free field zero mode $\varpi$ which has no
apparent relation to the generators of $U_q(sl(2))$ (and in
particular doesn't
seem to be connected with $J_3$ as one would expect from \cite{GS}),
is in fact part of the enveloping algebra. In the
$\psi$ basis $\varpi$ is shifted by the $\psi$ fields
(as $J_3$ by the $\xi$ fields). This suggests that there
may be another basis of generators $T_{\pm}$ and $T_3$ where
$T_3$ could be realized by $\varpi$ in the $\psi$ basis.
Let us show that this is in fact true at the formal level.
We are looking for an algebra of the type
$$
\Tpmo \Tto =q^{\mp 1} \Tto \Tpmo
$$
$$
\left [ \Tpo,\Tmo \right ]= F\left( \Tto \right)
$$
where F is a function to be determined.
Equation \ref{Ddef} suggests to write
\beq
\Do =\alpha_D ({\cal O}\left [q^{\beta T_3}\right ]+
{\cal O}\left [q^{-\beta T_3}\right ]),
\label{Dgendef}
\eeq
for some $\beta$. Furthermore, we will make the ansatz that there
exists a relation of the form
$$
\Jto = F_+\left( \Tto \right) \Tpo + F_-\left( \Tto \right)\Tmo
$$
with $F_\pm$ to be determined. Let us insert the above expressions
for $\Do$ and $\Jto$ into Eqs.\ref{consist1}, \ref{consist2}. {}From
Eq.\ref{consist1} we infer
$$
\alpha _D^2 = - {C_+ C_- \over (q-q^{-1})^2}, \quad \beta=\pm 1.
$$
Eq.\ref{consist2} implies\footnote{More precisely,
we can always transform the solution to this form by a suitable
redefinition
$T_\pm \to H_\pm(T_3)T_\pm$ with $H_+(T_3)H_-(T_3-1)=1$, which leaves
the algebra of $T_\pm$ invariant.}
$$
F_+\left( \Tto \right) = F_-\left( \Tto \right) =
{\alpha _3 \over (\Tto -\Tso)}
$$
and
\beq
F\left( \Tto \right) =
{\alpha_D (q-q^{-1})^2 \over \alpha^2_3 }
(\Tto -\Tso).
\label{Fform}
\eeq
$\Jpmo$ are then computable from formula  \ref{Dcom}. We finally get
$$
\Jpmo = {\alpha_\pm \over \Tto -\Tso} (\Tsto\Tpo+\Ttso\Tmo ),
$$
\beq
\Jto = {\alpha _3 \over (\Tto -\Tso)}
(\Tpo + \Tmo)
\label{JTrel}
\eeq
with
$$
\alpha_\pm ={\alpha_D \alpha_3 \over C_{\mp}}
$$
Choosing\footnote{In the case where $C_+C_-=0$, one finds that
$[\Tpo , \Tmo ]=0$, which gives a
contraction of $U_q(sl(2))$.}
\beq
  \alpha^2_3= \alpha_D (q-q^{-1})^2
( q^\demi -q^{-\demi})
\label{alprel}
\eeq
we obtain the standard commutation relation of
$U_{\sqrt{q}}(sl(2))$
$$
\Tpmo \Tto =q^{\mp 1} \Tto \Tpmo
$$
\beq
\left [ \Tpo,\Tmo \right ]= {\Tto - \Tso \over q^\demi -q^{-\demi}
}
\label{Tcomgen}
\eeq
 We go from
$U_q(sl(2))$ to $U_{\sqrt{q}}(sl(2))$ essentially because $F\left(
\Tto\right)$, the commutator of $\Tpo$ and $\Tmo$, is a linear
function
of $\Ttso$ according to Eqs. \ref{Dgendef} and \ref{Fform}.
 The representation
Eqs. \ref{JTrel} involves the inverse of the
operator
$\Tto -\Tso$ which is not defined for eigenstates of $T_3$ with
vanishing
eigenvalue. However, it turns out that the normalization of the
operators
$\Tpmo$ vanishes at this point as well, so that a well-defined
limit exists\footnote{Subtleties at $\varpi=0$ can actually be
expected for general
reasons, as it is the fixed point of the Weyl reflection symmetry
of
the theory \cite{BG}\cite{ANPS}.}.
\section{Transformation laws of the Bloch wave operators}
\label{bloch wave}
As alluded to already in the introduction, one a priori expects that
the ${\cal O}\! \left[J^a\right]$ operators generate the symmetries
of the Bloch wave basis as well. We will therefore
start by considering explicitly the action by braiding of the ${\cal
O}\! \left[J^a\right]$
on the Bloch wave fields.
\subsection{Action of the generators ${\cal O}\! \left[J^a\right]$
  on the Bloch waves.}
 In  previous works, the Bloch wave operators
have been handled   with  various  ($\varpi$ dependent)
normalizations\footnote{denoted by the
symbols $\psi$, $V$, $\widetilde V$ and $U$.}. It will be simplest to
 make use of  the $\psi$ fields of refs.\cite{GN,G1};
one may easily change to different normalizations afterwards.
 It is convenient to write the relation between $\psi$ and $\xi$
fields as
\beq
\xi_M^{(J)}(\sigma)=\sum_m\psi_m^{(J)}(\sigma) \> \U J, m,
M,\varpi, \quad
\psi_m^{(J)}(\sigma)=\sum_M\xi_M^{(J)}(\sigma) \V J, M, m,\varpi
\label{xipsi}
\eeq
On the other hand, for spin ${1\over 2}$ it is better  to let
\beq
\xi_\alpha^{({1\over 2})}=\sum_\beta u_{\alpha \beta}(\varpi)
\psi^{({1\over 2})}_\beta,
\quad
\psi^{({1\over 2})}_\alpha= \sum_\beta v_{\alpha \beta}(\varpi)
\xi^{({1\over 2})}_\beta
\label{udef}
\eeq
so that
$u_{\alpha \beta}(\varpi)=\U {{1\over
2}},\beta,\alpha,{\varpi+2\beta} .
$
Explicitly one has
\beq
u_{-\demi -\demi}=q^{(\varpi-1)/2},\quad
u_{-\demi \demi}=q^{-\varpi/2}, \quad
u_{\demi -\demi}=q^{-\varpi/2}, \quad
u_{\demi \demi}=q^{(\varpi+1)/2},
\label{uform}
\eeq
Greek indices take the values $\pm \demi$.
No summation over repeated indices is assumed  unless explicitly
indicated, in order to avoid
confusions. Since the fields $\xii_{\pm\demi} (\sigma_+)$
 generate the quantum group
transformations of the fields $\xi_M^{(J)}(\sigma)$, it is natural
to  study their action on
 the fields $\psi_m^{(J)}(\sigma)$. In order to relate new and old
structures,
let us start from the transformation laws of the $\xi$ fields
Eqs.\ref{act3+} and \ref{act+}, derived in ref.\cite{CGS1},  and
transform
$\xi_M^{(J)}$ into $\psi_m^{(J)}$ using Eq.\ref{xipsi}. One finds
at first
$$
\Jtop{\sigma_+} \psi_m^{(J)}(\sigma)=\sum_{n,\, M}
\psi_n^{(J)}(\sigma)\> \U J, n, M,\varpi q^M \Jtop{\sigma_+}
 \V J, M, m,\varpi ,
$$
$$
\Jpmop{\sigma_+} \psi_m^{(J)}(\sigma)=\sum_{n,\, N,\, M}
\psi_n^{(J)}(\sigma)\> \U J, n, N,\varpi (J_\pm)_{NM}
\Jtop{\sigma_+}
 \V J, M, m,\varpi +
$$
\beq
\sum_{n,\, M} \psi_n^{(J)}(\sigma)\> \U J, n, M,\varpi q^{-M}
\Jpmop{\sigma_+}
 \V J, M, m,\varpi  .
\label{xipsi1}
\eeq
Next, we have to braid the generators $\Jtop{\sigma_+} $ and
$\Jpmop{\sigma_+} $ with the
$\V J, M, m,\varpi $  coefficients.
This is done most simply in terms
of the $\psii$
fields, since they shift $\varpi$ in a simple way. Indeed, one has in
general
\beq
\psi_m^{(J)}\! (\sigma) \>\varpi= (\varpi+2m)\> \psi_m^{(J)}\!
(\sigma).
\label{shift}
\eeq
There are two cases. First, it follows from
Eq.\ref{Jopdef+} that $\Jtop{\sigma_+} $ and $\Jpop{\sigma_+} $ are
proportional to the
$\xii$ fields,
so that they can be immediately  reexpressed in terms of $\psii$
fields
using Eq.\ref{udef}. Second,
the expression of  $\Jmop{\sigma_+} $ in terms of the $\xii$ fields
is given by
Eq.\ref{monodr}, which makes use of the monodromy. Using
Eqs.\ref{udef}, \ref{uform},
one then derives the equation
\beq
\Jmop{\sigma_+} =\kappa_-^{(+)}\sum_\lambda u_{-\demi, \lambda}
q^{-(2\lambda \varpi+\demi)} \psii_{\lambda}.
\label{Jmpsi}
\eeq
 Using Eq.\ref{shift}, one deduces from
Eq.\ref{xipsi1} that
\beq
\Jtop {\sigma_+}  \psi_m^{(J)}(\sigma)=\kappa^{(+)}_3\,
\sum_{\lambda,\, n}
\psi_n^{(J)}(\sigma)\> \Jtpsi J,\lambda,n,m,\varpi \>
 u_{-\demi \lambda}(\varpi)\, \psii_\lambda(\sigma_+),
\label{psitr30}
\eeq
\beq
\Jpmop {\sigma_+}  \psi_m^{(J)}(\sigma)=\kappa^{(+)}_\pm \,
\sum_{\lambda,\, n}
\psi_n^{(J)}(\sigma)\> \Xpmpsi J,\lambda,n,m,\varpi \>
 u_{-\demi \lambda}(\varpi)\,
\psii_\lambda(\sigma_+)
\label{psitrp0}
\eeq
\beq
\Xpmpsi J,\lambda,n,m,\varpi= {1-q^{\pm 2}\over q^{\pm \demi}
}\Jpmpsi
J,\lambda,n,m,\varpi \>
+q^{\pm (2\lambda \varpi +\demi)}
 \Jspsi J,\lambda,n,m,\varpi .
\label{Xdef}
\eeq
These formulae, as well as  the equations  below, involve the
following
transforms  of the $U_q(sl(2))$ matrices
\beq
 \Jtspsi J,\lambda,n,m,\varpi=\sum_M  \U J,n,M,\varpi q^{\pm M}
\V J,M,m,{\varpi+2\lambda}
\label{Jtspsidef}
\eeq
\beq
 \Jpmpsi J,\lambda,n,m,\varpi=\sum_{N,\, M}\U J,n,N,\varpi \left
[J_\pm \right]_{NM}
\V J,M,m,{\varpi+2\lambda} ,
\label{Jpmpsidef}
\eeq
which will play a key role. At this point, the co-action
of our generators is not yet in a satisfactory form. We have to
rexpress the right hand sides in
terms of the generators. This is
straightforward, in principle
using Eq.\ref{udef}.
After that one may re-express the r.h.s. in terms of the generators
using Eqs.\ref{Jopdef+},
and/or Eq.\ref{monodr}. Now we meet two difficulties. First there is
an ambiguity, since
there are two $\xii$ fields for three generators\footnote{
There was no such difficulty for the action on the $\xi$ fields
because
there, the transformation matrices did not contain $\varpi$, so that
$\varpi$ - dependent linear combinations of $\xi$ fields as in
Eq.\ref{monodr}
 could be consistently viewed as separate operators.}.
Note that, although the
braiding-algebra  of the $\xi$ fields  does not involve $\varpi$,
Eq.\ref{monodr}
does contain $\cos (h\varpi)$.
Thus $\varpi$ belongs to the algebra in some way, and indeed it is
part of the operator
$\Dop{\sigma_+},{\sigma_-} $, as shown in Eq.\ref{Ddef}. Thus from
this viewpoint,
the transformation from $\xi_M^{(J)}$
to  $\psi_m^{(J)}$ (Eq.\ref{xipsi}) involves functions of the
generators, and the
$\psi_m^{(J)}$ fields
appear as complicated members of the enveloping algebra. This
explains the second difficulty,
namely that the formulae just derived  are not of the usual
co-action type. Of course one  may nevertheless rederive
the consequences of the algebra of the generators, following the line
of ref\cite{CGS1}.
  However, the resulting formulae are rather
involved, and the underlying symmetry structure of the Bloch wave
operator
algebra is difficult to extract in this way. Therefore we will pass
to a different form of the generators in the next subsection, where
this structure will become much more transparent, while still
equivalent
to the one generated by the ${\cal O}\! \left[J^a\right]$ operators.
On the way, it will be useful to collect some further formulae
characterizing
the action of the ${\cal O}\! \left[J^a\right]$ on the Bloch waves,
to which we will come back in section \ref{xitraforel}.

By performing a ($\varpi$ dependent) change of basis from the
structure
derived in ref.\cite{CGS1}, we have arrived at formulae where the
actions
of $\Jtop{\sigma_+}$ and
of $\Jpmop{\sigma_+}$ look  different.  Indeed,  the former involves
one matrix $\Jtpsi J,\lambda,n,m,\varpi$ while the latter involves
two, that is
$\Jpmpsi J,\lambda,n,m,\varpi$ and $\Jspsi J,\lambda,n,m,\varpi$. Our
next point is that
this  structure is redundant, and that the above actions may actually
all be described in terms of the matrix $\Jtpsi J,\lambda,n,m,\varpi$
alone.
As a first step, let us analyze our equations in terms of the
possible shifts
 of $\varpi$ between
bras and kets,
using Eqs.\ref{udef} and \ref{shift}. On the l.h.s. of
Eqs.\ref{psitrp0}
 and \ref{psitr30}, the possible shifts are $2m$ and $2m\pm 1$.
Thus the matrices $\Jtpsi J,\lambda,n,m,\varpi$ and
 $\Xpmpsi J,\lambda,n,m,\varpi$ are zero unless $n=m$ or $n=m\pm 1$.
 However, it can be seen easily that the
two matrices  $\Jpmpsi J,\lambda,n,m,\varpi$ and
 $\Jspsi J,\lambda,n,m,\varpi$, involved in Eq.\ref{Xdef}
do not share  this property. Thus many cancellations between them
take place,
and one should not consider them separately.   This, and the explicit
computation of the matrices,
  is achieved by re-doing the calculation otherwise.  We start from
the  braiding matrix
of $\psii$ with $\psi_m^{(J)}$ for general $J$ and $m$ derived in
ref.\cite{G1}.
  One has in general,
\beq
\psii_\alpha(\sigma) \,\psi_m^{(J)}(\sigma')=\sum_\beta \sum_n
S_{\epsilon  \phantom{(J)} \alpha m}^{(J)\, n\beta}(\varpi)
\, \psi_n^{(J)}(\sigma') \,\psii_\beta(\sigma),
\label{psibraid}
\eeq
where the non-vanishing matrix elements are given by ($\epsilon$ is
the
sign of $\sigma-\sigma'$ \footnote{for
$\sigma-\sigma' \in [-2\pi,2\pi]$, which
we consider here})
$$
S_{\epsilon  \phantom{(J)} -{1\over 2}\, m}^{(J)\, m\, -{1\over
2}}(\varpi)=
S_{\epsilon \phantom{(J)}  {1\over 2},\, -m}^{(J)\, -m,\, {1\over
2}}(-\varpi)=
{\lfloor \varpi+J+m \rfloor \over \lfloor \varpi \rfloor}
e^{ihm\epsilon}
$$
\beq
S_{\epsilon  \phantom{(J)} -{1\over 2},\, m}^{(J)\,m-1,\, {1\over
2}}(\varpi)=
{\lfloor J+m \rfloor \over \lfloor \varpi \rfloor}
e^{ih\epsilon (1-m-\varpi)}=
S_{\epsilon \phantom{(J)}  {1\over 2},\, -m}^{(J)\, -m+1,\, -{1\over
2}}(-\varpi).
\label{Sdef}
\eeq
Here,
$$
\lfloor x\rfloor :={q^x-q^{-x}\over q-q^{-1}}
$$
denotes $q$ - numbers.
Making use of  Eqs.\ref{udef}, one deduces from
Eq.\ref{psibraid} that
$$
\xii_{\gamma}(\sigma_\pm)\>
\psi_m^{(J)}(\sigma)=\sum_{\lambda,\,\alpha,\,  n}
\psi_n^{(J)}(\sigma) u_{\gamma \, \alpha}(\varpi-2n) S_{\pm
\phantom{(J)} \alpha m}^{(J)\,
n\lambda}(\varpi-2n) \psii_\lambda(\sigma_\pm).
$$
Comparing with Eq.\ref{psitr30}, \ref{psitrp0},  one infers  that
$$
 \Jtpsi J,\lambda,n,m,\varpi=
{u_{\mp { 1\over 2},\, n-m+\lambda}(\varpi-2n)\over u_{\mp{1\over
2},\,
\lambda}(\varpi)} S_{\pm   \>  n-m+\lambda,\,  m}^{(J)\,
n\, \lambda}(\varpi-2n)
$$
\beq
 \Xpmpsi J,\lambda,n,m,\varpi=
{u_{\pm { 1\over 2},\, n-m+\lambda}(\varpi-2n)\over u_{\mp{1\over
2},\,
\lambda}(\varpi)} S_{\pm   \>  n-m+\lambda,\,  m}^{(J)\,
n\, \lambda}(\varpi-2n).
\label{explma}
\eeq
Notice that both choices of the signs on the right hand sides lead to
the same
result.
The equalities for the lower signs are easily verified using the
monodromy
properties of the $\xi^{(\demi)}_{-\demi}$ field \cite{CGS1}:
\beqa
\xii_{-\demi}(\sigma+2\pi)&=& \xii_{\demi}(\sigma), \nnn
\xii_{-\demi}(\sigma-2\pi)&=&2q^{-\demi}
\cos(h\varpi) \> \xii_{-\demi}(\sigma)
-q^{-1}\xii_{\demi}(\sigma)\nnn
\label{ximonodr}
\eeqa
which allow to identify $\xi^{(\demi)}_\demi(\sigma_-)$ with
$\xi^{(\demi)}_{-\demi}(\sigma_+)$ for the first equation
\ref{explma},
and $\Jmop{\sigma_+}$ with $\xi^{(\demi)}_{-\demi}(\sigma_-)$ for the
second.
Eqs.\ref{ximonodr}
are derived from the simple monodromy behaviour of the
Bloch waves:
\beq
\psii_{\alpha}(\sigma+2\pi)= q^{ 2\alpha \varpi+ \demi}
\psii_{\alpha}(\sigma)
\label{monodpsi}
\eeq
It follows from the formulae just given that the only nonzero matrix
elements of
$\Jtpsi
J,\lambda,n,m,\varpi$  are
\beq
\Jtpsi J,{-\demi},n,n,\varpi=
{\lfloor \varpi+J-n\rfloor \over \lfloor \varpi-2n\rfloor},\quad
\Jtpsi J,{-\demi},n,{n-1},\varpi=-q^{-{1\over 2}}
{\lfloor J-n+1\rfloor \over \lfloor \varpi-2n\rfloor},
\label{je1}
\eeq
\beq
\Jtpsi J,{\demi},n,n,\varpi=
{\lfloor \varpi-J-n\rfloor \over \lfloor \varpi-2n\rfloor},\quad
\Jtpsi J,{\demi},n,{n+1},\varpi=q^{{1\over 2}}
{\lfloor J+n+1\rfloor \over \lfloor \varpi-2n\rfloor}.
\label{je2}
\eeq
Using   the explicit expressions Eq.\ref{uform},
one sees that
\beq
\Xpmpsi J,\lambda,n,m,\varpi = q^{\pm
2(n+\lambda-m)(\varpi-2n)\pm \demi }
\Jtpsi J,\lambda,n,m,\varpi
\label{explma2}.
\eeq
from which it follows since $n-m+\lambda =\pm \demi$
\beq
q^{-\demi} \Xppsi J,\mu,n,p,\varpi +
q^{\demi} \Xmpsi J,\mu,n,p,{\varpi} =\sum _m  D(\varpi)_{nm}
 \Jtpsi J,\mu,m,p,\varpi
\label{calc5}
\eeq
\beq
q^{\demi} \Xppsi J,\mu,n,p,\varpi +
q^{-\demi} \Xmpsi J,\mu,n,p,{\varpi} = \sum _m
 \Jtpsi J,\mu,n,m,\varpi \> D(\varpi +2\mu )_{mp}.
\label{calc7}
\eeq
where
\beq
D(\varpi)_{nm} =\delta_{n,\, m}
(q^{\varpi-2n}+q^{-\varpi+2n}).
\label{Ddef2}
\eeq
Returning to the transformation laws of the $\psi^{(J)}$ fields, one
sees
that we may finally 	rewrite them under the form
$$
{1\over \kappa^{(+)}_3} \Jtop{\sigma_+} \psi^{(J)}_m(\sigma)=
\sum_{\lambda ,n} \psi^{(J)}_n(\sigma)
\Jtpsi J,{m-n+\lambda},n,m,\varpi  u_{-\demi ,m-n+\lambda}
\psi^{(\demi)}_{m-n+\lambda}(\sigma_+),
$$
\beq
{1\over  \kappa^{(+)}_\pm}
\Jpmop{\sigma_+} \psi^{(J)}_m(\sigma)=
 \sum_{\lambda ,n}  q^{\pm (2\lambda \varpi+\demi)}
\psi^{(J)}_n(\sigma)
\Jtpsi J,{m-n+\lambda},n,m,\varpi  u_{-\demi ,m-n+\lambda}
\psi^{(\demi)}_{m-n+\lambda}(\sigma_+).
\label{transf2}
\eeq
Up to the factors $q^{\pm (\demi+2\lambda \varpi)}$, the right-hand
sides
involve the same matrices $\Jtpsi J,{m-n+\lambda},n,m,\varpi  $
and operators $u_{-\demi ,m-n+\lambda}
\psi^{(\demi)}_{m-n+\lambda}(\sigma_+)$.
\subsection{Changing basis on the generators}
\label{basischange}
Up to now we have performed a change of basis only on the fields but
not the
generators. However, as announced above, the most natural realization
of the quantum group on the
Bloch wave fields turns out to be given in terms of new generators
which
are represented by Bloch waves rather than $\xi$ fields. More
precisely,
the raising
and lowering generators will be given in terms of the
$\psi^{(\demi)}_\alpha$,
 while the Cartan generator will be represented by the zero mode
$\varpi$.
Compared to the last subsection, where it appeared to be merely a
parameter
appearing in
the basis transformation between the $\xi$ and the $\psi$ fields, the
role
of the zero mode thus changes drastically. In this way, we will
arrive  at a
FP realization of the transformation
Eq.\ref{JTrel} by  generators
 denoted $\Ttop{\sigma_+}$, and
$\Tpmop{\sigma_+}$. In the new picture, functions of $\varpi$ which
appear
in the coproduct action of the generators on the Bloch waves are much
easier
to handle because the commutation relations of functions of $T_3$
with the other generators are very simple.
We first observe that it is just the transformation between $\xii$
and
$\psi^{(\demi)}$ fields (Eqs.\ref{udef}) which may be  used to
realize a redefinition of generators similar to Eqs.\ref{JTrel}
  of section \ref{covbasis}. The steps of the derivation  are as
follows.
 First  according to Eq.\ref{shift}
 $\psi_m^{(J)}$ shift
$\varpi$ by $2m$,
so that we will  realize the first line of Eqs.\ref{Tcomgen}, if we
let
\beq
\Ttsop{\sigma_+}=q^{\mp \varpi}.
\label{Ttsopdef}
\eeq
and choose $\Tpmop{\sigma_+}$ to be proportional to $\psii_{\pm
\demi}$,
as suggested above.
Second, the first relation of Eq.\ref{udef} with $\alpha=-\demi$
may be rewritten as
\beq
\Jtop{\sigma_+} =\kappa_3^{(+)} \left ( \Ttop{\sigma_+}-
\Tsop{\sigma_+}\right)^{-1} \left [ \Tpop{\sigma_+}+
\Tmop{\sigma_+}\right ],
\label{Toprel1}
\eeq
if we let
\beq
\Tpmop{\sigma_+} =(q^{-\varpi}-q^\varpi)
u_{-\demi, \pm \demi}(\varpi) \psii_{\pm \demi}(\sigma_+),
\label{Topdef}
\eeq
Third, the first relation of Eq.\ref{udef} with $\alpha=\demi$
may be transformed into
\beq
\Jpop{\sigma_+} =\kappa_+^{(+)} q^{\demi}
\left ( \Ttop{\sigma_+}-
\Tsop{\sigma_+}\right)^{-1} \left [ \Tsop{\sigma_+} \Tpop{\sigma_+}+
\Ttop{\sigma_+} \Tmop{\sigma_+}\right ].
\label{Toprel3}
\eeq
Fourth, Eq.\ref{Jmpsi} gives
\beq
\Jmop{\sigma_+} =\kappa_-^{(+)} q^{-\demi}
\left ( \Ttop{\sigma_+}-
\Tsop{\sigma_+}\right)^{-1} \left [ \Ttop{\sigma_+}
\Tpop{\sigma_+}+
\Tsop{\sigma_+} \Tmop{\sigma_+}\right ].
\label{Toprel2}
\eeq
Fifth, Eq.\ref{Ddef} may be rewritten as
\beq
\Dop{\sigma_+},{\sigma_+'}  =
{C_+(\sigma_+,\sigma_+') \over q-q^{-1}}
\left [ \Ttop{\sigma_+}
+\Tsop{\sigma_+}  \right ].
\label{Dopdef}
\eeq
One sees that the FP version of the general Eqs.\ref{JTrel}
 is realized with
$$
\alpha_\pm=q^{\pm \demi} \kappa_\pm^{(+)},\quad
\alpha_3=\kappa_3^{(+)},
$$
\beq
\alpha_D={C_+(\sigma_+,\sigma_+')\over q-q^{-1}}
\label{alphaop}
\eeq
and with this choice, we arrive at the FP relations
$$
\Tpmop{\sigma_+} \Ttop{\sigma_+'}= q^{\mp 1}
\Ttop{\sigma_+'} \Tpmop{\sigma_+}
$$
\beq
\Tpop{\sigma_+}  \Tmop{\sigma_+'} -
\Tmop{\sigma_+}   \Tpop{\sigma_+'} =
{C_+(\sigma_+,\sigma_+')\over q^{\demi} \kappa_+^{(+)}
\kappa_3^{(+)} } \left [  \Ttop{\sigma_+} - \Tsop{\sigma_+}
\right ].
\label{FPcom}
\eeq
The second equation can be verified, for instance, using
Eqs.\ref{Cplus},
\ref{Topdef} and the explicit expression for
$\xi^{[\demi , \demi](0)}_0(\sigma_+, \sigma'_+)$ in terms of the
$\psi$ fields (see Eq. (5.3)
of ref.\cite{CGS1}).
 Up to a multiplicative factor, this is the FP realization of the
$U_{\sqrt{q}}(sl(2))$
commutation relation of section 2. Note that this multiplicative
factor is itself
proportional to the field $C_+\propto \xic (\sigma_+,\sigma_+')$, so
that it is not
really possible to get rid of it by changing the normalization of
$\Tpmop  {\sigma_+} $ as
was done in section 2 by imposing Eq.\ref{alprel}.
Altogether the present FP realization of the commutation relation is
such
that $\Ttsop{\sigma_+} $, $\Tpmop{\sigma_+} $ have grading zero and
one, respectively.
The fact that   $\Ttsop{\sigma_+} $ has grading zero is of course
necessary to
be able to define the right hand side of Eqs.\ref{Toprel3} and
\ref{Toprel2}.
 Next, one substitutes  Eq.\ref{Toprel1}
into the first line of Eq.\ref{transf2}. Making use of
Eq.\ref{shift}, one
sees that  $\Tpmop {\sigma_+} $ give a total shift of
$\varpi$ of $m\pm 1/2$ respectively  on the
left hand side.  Separating the corresponding contributions on the
right hand side
 one   obtains
\beq
\Tpmop{\sigma_+} \psi_m^{(J)}(\sigma)= \sum_n\psi_n^{(J)}(\sigma) \>
\Tpsi J,{\pm \demi},n,m,\varpi \> \Top{+},{(m-n \pm
\demi)},{\sigma_+},
\label{Taction}
\eeq
\beq
\Tpsi J,\lambda,n,m,\varpi=
{\sin [h(\varpi-2n)]\over \sin [h(\varpi)]}\>
\Jtpsi J,m-n+\lambda,n,m,\varpi .
\label{Tpsidef}
\eeq
This is consistent with the definition Eq.\ref{Topdef} since by
construction,
$ \Tpsi J,\lambda,n,m,\varpi$ vanishes unless $m-n+\lambda=\pm
\demi$.
On the right hand side of the first equation,
the index of $T$ is always $\pm 1$, and
from now on we  use both notations $\Tpmop{\sigma_+}$
and ${\cal O}\! \left[T_{\pm 1}\right]_{\sigma_+}$.
Furthermore, let us note here that the FP braiding relations of our
generators lead to the following relations
on $\Tpsi J,\lambda,n,m,\varpi $:
\beqa
\Tpsi J,\lambda,m,m,\varpi \> \Tpsi J,-\lambda,m,m,{\varpi+2\lambda}&
- &
\Tpsi J,{-\lambda},m,\, m+2\lambda ,\varpi \> \Tpsi
J,\lambda,m+2\lambda\, ,m,{\varpi+2\lambda}
=
{\lfloor \varpi-2m \rfloor\over \lfloor \varpi \rfloor}\nnn
\Tpsi J,\lambda,m+2\lambda\, ,m,\varpi \> \Tpsi
J,-\lambda,m,m,{\varpi-2\lambda}& - &
\Tpsi J,{-\lambda},m+2\lambda,\, m+2\lambda ,\varpi \>
\Tpsi J,\lambda,m+2\lambda\, ,m,{\varpi-2\lambda}=0.
\label{mess}
\eeqa
Making use of the relation  to the original matrix realization of
$U_q(sl(2))$ displayed
in this section, one may rederive these relations.
This will be  done later on.
 Note that
these relations involve particular values of the indices $m$ and $n$
of
 $\Tpsi J,{\lambda},m,n,\varpi  $ {\bf without summation over these
indices}.
So far, we have used $\psi$ fields at point $\sigma _+$ to realize
the operators $\Tpmop{\sigma_+} $. We can do exactly the same
thing with $\psi$ fields at point $\sigma _-$, they will realize
operators $\Tpmop{\sigma_-}$. As in the case of the $\xi$ fields
both representations are related by monodromy which is diagonal
for the $\psi$ fields (see Eq. \ref{monodpsi}).
More precisely we have
\beqa
\Tpmop{\sigma_-}& =&(q^{-\varpi}-q^\varpi)
u_{\demi, \pm \demi}(\varpi) \psii_{\pm \demi}(\sigma_-)
 =(q^{-\varpi}-q^\varpi)
u_{-\demi, \pm \demi}(\varpi)
\psii_{\pm \demi}(\sigma_- + 2\pi) \nnn
&=& \Tpmop{\sigma'_+} \>
\hbox{ at} \> \sigma'_+=\sigma_-+2\pi .
\label{Topdef-}
\eeqa
Next we turn to the action of $\Top{+},{\lambda},{\sigma_+}$ on a
product of $\psi$ fields, which has a very simple structure
in terms of the $ \Tpsi J,\alpha,n,m,\varpi$ matrices.
 One has
$$
\Top{+},{\lambda},{\sigma_+}
 \psi_{m_1}^{(J_1)}(\sigma_1) \psi_{m_2}^{(J_2)}(\sigma_2)=
$$
\beq
\sum_{n_1, \, n_2}
 \psi_{n_1}^{(J_1)}(\sigma_1) \psi_{n_2}^{(J_2)}(\sigma_2)\>
\LTpsi J_1,J_2,\lambda,n_1,n_2,m_1,m_2,\varpi \>
\Top{+},{(m_1+m_2-n_1-n_2+\lambda)},{\sigma_+},
\label{Tcopract}
\eeq
where
\beq
\LTpsi J_1,J_2,\lambda,n_1,n_2,m_1,m_2,\varpi =
\Tpsi J_1,\lambda,n_1,m_1,{\varpi-2n_2} \> \Tpsi
J_2,{m_1-n_1+\lambda},n_2,m_2,\varpi .
\label{Tcopmat}
\eeq
Moreover, we may rewrite Eq.\ref{Taction} under the following  form
similar to the general
co-product action\cite{MaSc} discussed in ref.\cite{CGR1}
\beq
\Top{+},{\lambda},{\sigma_+} \psi_m^{(J)}(\sigma)= \sum_n
\psi_n^{(J)}(\sigma) \>
\Lambda\left(\Top{+},{\lambda},{\sigma_+}\right)_{nm}
\label{Taction2}
\eeq
where
\beq
\Lambda\left(\Top{+},{\lambda},{\sigma_+} \right)_{nm}=
\Tpsi J,\lambda,n,m,\varpi \> \Top{+},{(m-n+\lambda)},{\sigma_+}
\label{Tcopmop}
\eeq
Clearly Eq.\ref{Tcopmop} is analogous  to Eq.\ref{Tcopmat}, the
second matrix being replaced
by the generator.
Notice however that  the second term depends upon the
row/column indices of the first matrix, so that a priori it doesn't
seem to have an interpretation in terms of a coproduct.  We will
see in  section 4 that such an interpretation in fact becomes
available once we realize that the true underlying symmetry is not
$U_{\sqrt q}(sl(2))$ but $U_q(sl(2))\otimes U_q(sl(2))$.
\section{Hidden $U_q(sl(2)) \otimes U_q(sl(2))$ structure.}
\subsection{The matrix algebra}
In order to retransform Eqs.\ref{mess} into matrix
relations, it is convenient
to introduce  two  sets of matrices $A^\lambda$ and $B^\lambda$
defined by
\beq
\A J,{\lambda},n,m,\varpi = \Tpsi J,{\lambda},m,m,\varpi \>
\delta_{m,n}
\qquad
\B J,{\lambda},n,m,\varpi = \Tpsi J,{ - \lambda},{m -2\lambda
},m,\varpi
\>
\delta_{n ,m-2\lambda }.
\label{defAB}
\eeq
Eqs.\ref{mess} become
$$
\sum _p \left(
\B J,{\lambda},n,p,{\varpi} \>
\A J,{\lambda},p,m,{\varpi +2\lambda} -
\A J,{\lambda},n,p,{\varpi} \>
\B J,{\lambda},p,m,{\varpi +2\lambda} \right) = 0
$$
\beq
\sum _p \left(
\A J,{\lambda},n,p,{\varpi} \> \A J,{-\lambda},p,m,{\varpi
+2\lambda} -
\B J,{\lambda},n,p,{\varpi} \> \B J,{-\lambda},p
,m,{\varpi+2\lambda} \right) =
{\lfloor \omega -2m \rfloor \over  \lfloor \omega \rfloor}
 \delta _{mn}.
\label{mess2}
\eeq
Eq.\ref{Taction} takes the form
\beq
\Top{+},{\lambda},{\sigma_+} \psi_m^{(J)}(\sigma)=
\sum_n\psi_n^{(J)}(\sigma) \>
\left \{\A J,{\lambda},n,m,{\varpi} \> \Top{+},{\lambda},{\sigma_+}
+ \B J,{-\lambda},n,m,{\varpi} \>
\Top{-},{\lambda},{\sigma_+}\right\}
\label{TactAB}
\eeq
There remains to handle the $\varpi$ dependence which prevents
Eqs.\ref{mess2}
from being ordinary
commutation relations.   For this, we interpret $\varpi$ as an
additional
index, by writing
\beq
\psi^{(J)}_{\rho,\varpi}=\psi^{(J)}_m P_{\varpi}, \qquad
\rho = \varpi -2m
\label{MS}
\eeq
where $P_{\varpi}$ is the projector onto the Verma module
characterized
by\footnote{Of course, this is nothing but a slightly different
notation
for the standard chiral vertex operators $V^I_{JK}$ of conformal
field theory
-cf. ref. \cite{CGR1}.}$\varpi$. Since
$$
\Top{+},\lambda,{\sigma_+} P_{\varpi}=
P_{\varpi-2\lambda}\Top{+},\lambda,{\sigma_+}
$$
Eqs.\ref{TactAB} become
\beq
\Top{+},\lambda,{\sigma_+}\psi^{(J)}_{\rho ,\varpi}  (\sigma)=
\sum_{\rho' , \varpi'}
 \psi^{(J)}_ {\rho' ,\varpi'}  (\sigma) \left \{
[A^\lambda]^{(J)}_{\rho'\varpi',\rho\varpi} \Top{+},
\lambda,{\sigma_+}
 +
[B^{-\lambda}]^{(J)}_{\rho'\varpi',\rho\varpi}
\Top {-},{\lambda},{\sigma_+}\right \}
\label{2TactAB}
\eeq
where we have defined
$$
[A^\lambda]^{(J)}_{\rho'\varpi',\rho\varpi}= \A
J,\lambda,{{\varpi'-\rho'\over 2}},
{{\varpi-\rho\over 2}},{\varpi'}
\delta_{\varpi',\varpi-2\lambda}
$$
\beq
[B^\lambda]^{(J)}_{\rho'\varpi',\rho\varpi}=\B
J,\lambda,{{\varpi'-\rho'\over 2}},
{{\varpi-\rho\over 2}},{\varpi'}   \delta_{\varpi',\varpi-2\lambda}.
\label{newAB}
\eeq
At this point it is convenient to introduce
\beq
[q^{\Omega_R}]_{\rho'\varpi',\rho\varpi}= q^{\varpi} \delta_{\rho'
,\rho}
\delta_{\varpi' ,\varpi},  \quad
[q^{\Omega_L}]_{\rho'\varpi',\rho\varpi}= q^{\rho} \delta_{\rho'
,\rho}
\delta_{\varpi' ,\varpi}.
\label{Omegdef}
\eeq
Eqs.\ref{mess2} become true matrix  relations
\beqa
A^{\lambda} A^{-\lambda}- B^{\lambda} B^{-\lambda}&=&
{\lfloor \Omega_L\rfloor \over \lfloor \Omega_R\rfloor}, \nnn
\left [ A^{\lambda},  B^{\lambda}\right ]&=&0.
\label{mess3}
\eeqa
Making use of Eqs.\ref{je1}, \ref{je2}, \ref{Tpsidef}, \ref{defAB}
one derives
the explicit expressions
$$
[A^\lambda]^{(J)}_{\rho'\varpi',\rho\varpi}=
{\lfloor (\rho'+\varpi')/2-2\lambda J \rfloor \over \lfloor \varpi'
\rfloor}
\delta_{(\rho'+\varpi')/2 ,(\rho+\varpi)/2-2\lambda}
\> \delta_{(\rho'-\varpi')/2 ,(\rho-\varpi)/2}
$$
\beq
[{}B^{\lambda}{}]^{(J)}_{\rho'\varpi',\rho\varpi}=
{q^{\lambda}
\lfloor (-\rho+\varpi)/2+2\lambda J \rfloor \over \lfloor \varpi'
\rfloor}
\delta_{(-\rho'+\varpi')/2 ,(-\rho+\varpi)/2-2\lambda}
\> \delta_{(\rho'+\varpi')/2 ,(\rho+\varpi)/2}.
\label{expAB}
\eeq
These last  formulae suggest to define
\beqa
\CAn J,{2\lambda},{\rho'},{\varpi'},\rho,\varpi &=&
2\lambda
\An J,{-\lambda},\rho',\varpi',\rho,{\varpi} \> \lfloor \varpi'
\rfloor
\qquad
\CA J,  {\rho'},{\varpi'},\rho,\varpi = q^{\varpi+\rho\over 2} \delta
_{\rho\rho'}\delta_{\varpi\varpi'}
\nnn
\CBn J,{2\lambda},{\rho'},{\varpi'},\rho,\varpi &=&
2\lambda
\Bn J,{-\lambda},{\rho'},{\varpi'},\rho,{\varpi} \> \lfloor \varpi'
\rfloor
\qquad
\CB J,  {\rho'},{\varpi'}, \rho,\varpi =
q^{\varpi-\rho \over 2} \delta _{\rho\rho'}
\delta_{\varpi\varpi'}
\label{defCAB}
\eeqa
We use  the shorthand notation  ${\bf{\cal A}}_\pm$,
${\bf{\cal B}}_\pm$
for the matrices $\CAn J,{2\lambda},{\rho'},{\varpi'},\rho,\varpi $,
$\CBn J,{2\lambda},{\rho'},{\varpi'},\rho,\varpi $ and similarly
for the diagonal generators. Writing down  the explicit expressions
derived from Eqs.\ref{je1} \ref{je2} \ref{Tpsidef} \ref{newAB},
one sees that, in the spaces where they act non-trivially,
 these matrices are in fact equal to the  ones of the standard
realization of $U_q(sl(2))$, up to simple changes of normalization.
 Accordingly, the following
matrix algebra holds:
\beq
\Bigl [{\bf{\cal A}}_+,\>  {\bf{\cal A}}_- \Bigr ]  =\lfloor 2
{\bf{\cal A}}_3
\rfloor, \qquad
{\bf{\cal A}}_\pm q^{{\bf{\cal A}}_3}=q^{\mp 1}q^{{\bf{\cal A}}_3}
{\bf{\cal A}}_\pm;
\label{CAA}
\eeq
\beq
\Bigl [{\bf{\cal B}}_+,\>  {\bf{\cal B}}_- \Bigr ]  =\lfloor 2
{\bf{\cal B}}_3
\rfloor, \qquad
{\bf{\cal B}}_\pm q^{{\bf{\cal B}}_3} =q^{\mp 1}q^{{\bf{\cal B}}_3}
{\bf{\cal B}}_\pm
\label{CBB}
\eeq
\beq
{\bf{\cal B}}_a {\bf{\cal A}}_b
 -
{\bf{\cal A}}_b{\bf{\cal B}}_a = 0
\label{ComAB}
\eeq
\beq
{\bf{\cal A}}_{-}{\bf{\cal A}}_{+}
+\left (\lfloor {\bf{\cal A}}_3+1/2  \rfloor \right )^2 =
{\bf{\cal B}}_{-}{\bf{\cal B}}_{+}
+\left (\lfloor {\bf{\cal B}}_3 +1/2 \rfloor \right )^2
\label{relCAB}
\eeq
The fact that the matrices ${\cal A}$ and ${\cal B}$ commute is
obvious
since they act non-trivially in spaces which are orthogonal.
The last equation   just imposes that the
 Casimir operators associated to the
two  $U_q(sl(2))$ be equal for this matrix representation, and be
given by
$ \left (\lfloor J+1/2  \rfloor \right )^2$. Effectively, this means
that
what we are dealing with is  not truly $U_q(sl(2)) \otimes
U_q(sl(2))$
but a reduction of it to five generators.

Finally, let us return to the action on a product of fields
already displayed on Eqs.\ref{Tcopract}, \ref{Tcopmat},
in order to understand
better the co-product. For this we rewrite
the product of $\psi$ fields using the notation introduced by
Eq.\ref{MS}.
Clearly
\beq
\psi^{(J_1)}_{m_1} \psi^{(J_2)}_{m_2} P_{\varpi_2}=
\psi^{(J_1)}_{\rho_1, \varpi_1} \psi^{(J_2)}_{\varpi_1, \varpi_2},
 \qquad
\rho_1 = \varpi_2 - 2(m_1+m_2), \> \varpi_1=\varpi_2-m_2.
\label{matching}
\eeq
The existence of a single value of the zero mode intermediate state
is what restricts
 us to only consider the product just written instead of the true
zero-mode tensor product
$\psi^{(J_1)}_{\rho_1, \varpi_1} \psi^{(J_2)}_{\rho_2, \varpi_2}$.
The corresponding
 matching condition,
\beq
\varpi_1=\rho_2,
\label{mcdt}
\eeq
 will play a key role.
Now we apply Eq.\ref{2TactAB} twice obtaining
$$
\Top{+},\lambda,{\sigma_+}\psi^{(J_1)}_{\rho_1 ,\varpi_1}  (\sigma_1)
\psi^{(J_2)}_{\rho_2 ,\varpi_2}  (\sigma_2)\delta_{\varpi_1,\rho_2}=
\sum_{\rho_1', \rho_2',\varpi'_1, \varpi_2'}
\psi^{(J_1)}_{\rho'_1 ,\varpi'_1}  (\sigma_1)
\psi^{(J_2)}_{\rho'_2 ,\varpi'_2}  (\sigma_2)\delta_{\varpi_1,\rho_2}
\times
$$
\beq
 \left \{
\tilde\Lambda\left
( A^\lambda\right)^{(J_1,J_2)}
_{\rho_1', \varpi_1',\rho'_2, \varpi_2';\,  \rho_1, \varpi_1,
\rho_2,\varpi_2 }
\Top{+}, \lambda,{\sigma_+}\right.
 +
\left.\tilde\Lambda\left
( B^{-\lambda}\right)^{(J_1,J_2)}
_{\rho_1', \varpi_1', \rho'_2\varpi_2';\,  \rho_1, \varpi_1,
\rho_2,\varpi_2 }
\Top{-}, \lambda,{\sigma_+}\right \}
\label{3TactAB}
\eeq
where
\begin{eqnarray}
\lefteqn{\tilde\Lambda\left
( A^\lambda\right)^{(J_1,J_2)}
_{\rho_1', \varpi_1', \rho_2',\varpi_2';\,  \rho_1, \varpi_1,
\rho_2,\varpi_2 }=}
\nnn
&&[A^\lambda]^{(J_1)}
_{\rho_1', \varpi_1',\,  \rho_1, \varpi_1}
 [A^\lambda]^{(J_2)}
_{\rho_2', \varpi_2',\, \rho_2, \varpi_2 }
+
[B^{-\lambda}]^{(J_1)}
_{\rho_1', \varpi_1',\,  \rho_1, \varpi_1}
 [B^\lambda]^{(J_2)}
_{\rho_2', \varpi_2',\, \rho_2, \varpi_2 }
\label{copA} \\
\lefteqn{ \tilde\Lambda\left
( B^\lambda\right)^{(J_1,J_2)}
_{\rho_1', \varpi_1', \rho_2',\varpi_2';\,  \rho_1, \varpi_1,
\rho_2,\varpi_2 }=}
\nnn
&&[A^{-\lambda}]^{(J_1)}
_{\rho_1', \varpi_1',\,  \rho_1, \varpi_1}
 [B^\lambda]^{(J_2)}
_{\rho_2', \varpi_2',\, \rho_2, \varpi_2 }
+
[B^{\lambda}]^{(J_1)}
_{\rho_1', \varpi_1',\,  \rho_1, \varpi_1}
 [A^\lambda]^{(J_2)}
_{\rho_2', \varpi_2',\, \rho_2, \varpi_2 }
\label{copB}
\end{eqnarray}
The interpretation of $\Omega_L$ and $\Omega_R$ in terms of Verma
modules
suggests to define
\beq
\tilde \Lambda(q^{\Omega_L})=q^{\Omega_L}\otimes {\bf 1},
\quad \tilde \Lambda(q^{\Omega_R})={\bf 1}\otimes q^{\Omega_R}
\label{copom}
\eeq
where the obvious index structure has been suppressed.
The mapping $\tilde\Lambda$
cannot yet be interpreted as a coproduct. This is because the algebra
Eqs. \ref{CAA} - \ref{ComAB} turns out to be preserved by $\tilde
\Lambda$
only on a subspace of the tensor product of the representation
spaces.
The necessary projection is defined by the matching condition Eq.
\ref{mcdt}.
 The important point here is that the structure of the $[A^\lambda]$
and
$[B^\lambda]$ is precisely such that $\tilde \Lambda$ respects
the matching condition, i.e.
$$
\tilde\Lambda\left
( A^\lambda\right)^{(J_1,J_2)}
_{\rho_1', \varpi_1', \rho_2'\varpi_2';\,  \rho_1, \varpi_1,
\rho_2,\varpi_2 }
\delta_{\varpi_1,\rho_2}=
\delta_{\varpi_1',\rho_2'}
\tilde\Lambda\left
( A^\lambda\right)^{(J_1,J_2)}
_{\rho_1', \varpi_1', \rho_2'\varpi_2';\,  \rho_1, \varpi_1,
\rho_2,\varpi_2 }
$$
and likewise for $\tilde\Lambda\left(B^\lambda\right)$.
Therefore we introduce the restriction
$$
\Lambda\left(A^\lambda\right)\equiv
\tilde\Lambda\left(A^\lambda\right)
P=P\tilde \Lambda \left(A^\lambda\right)
$$
of $\tilde \Lambda$ to the subspace defined by Eq. \ref{mcdt},
with $P$ the corresponding projector.
The above equations are then
 written compactly as
$$
\Lambda (A^{\lambda}) = (A^\lambda \otimes
A^\lambda+B^{-\lambda}\otimes
B^\lambda) P \equiv
A^{\lambda} \underbrace \otimes A^{\lambda} +
B^{-\lambda}  \underbrace \otimes B^{\lambda}
$$
\beq
\Lambda (B^{\lambda}) = (A^{-\lambda}\otimes B^\lambda+B^\lambda
\otimes
A^\lambda) P\equiv
A^{-\lambda} \underbrace  \otimes B^{\lambda} +
B^{\lambda}  \underbrace \otimes A^{\lambda}
\label{coprodAB}
\eeq
and
\beq
\Lambda(q^{\Omega_R})={\bf 1}\underbrace\otimes q^{\Omega_R} \qquad
 \Lambda(q^\Omega_L)= q^\Omega_L \underbrace \otimes {\bf 1}
\label{coprodom}
\eeq
where the "braced" tensor product serves as a convenient shorthand
notation for
the projection.
At the abstract level,
without refering to a matrix realization, the present definition of
$\underbrace \otimes$ is supposed to satisfy
\beq
q^{\Omega_R}
\underbrace \otimes {\bf 1}=
{\bf 1}  \underbrace \otimes
q^{\Omega_L}
\label{match}
\eeq
 In order to show
that the coproduct just defined respects the matrix relations
Eqs.\ref{CAA}--\ref{relCAB}, one   rewrites  it in terms
of the generators of
 $U_q(sl(2)) \otimes U_q(sl(2))$:
$$
\Lambda ({\bf{\cal A}} _{\pm}) =
{\pm 1 \over \lfloor {\bf{\cal A}}_3+ {\bf{\cal B}}_3\rfloor}
{\bf{\cal A}} _{\pm }
 \underbrace \otimes {\bf{\cal A}} _{\pm}
+
 {\mp 1 \over \lfloor {\bf{\cal A}}_3+ {\bf{\cal B}}_3\rfloor}
{\bf{\cal B}} _{\mp}
 \underbrace \otimes {\bf{\cal B}} _{\pm}
$$
\beq
\Lambda ({\bf{\cal B}} _{\pm}) =
{\mp 1  \over \lfloor {\bf{\cal A}}_3+ {\bf{\cal B}}_3\rfloor}
{\bf{\cal A}} _{\mp }
 \underbrace  \otimes {\bf{\cal B}} _{\pm    }
+
{\pm 1 \over \lfloor {\bf{\cal A}}_3+ {\bf{\cal B}}_3\rfloor}
{\bf{\cal B}} _{\pm}
 \underbrace \otimes {\bf{\cal A}} _{\pm}.
\label{coprodCAB}
\eeq
Note that, due to the property \ref{match}, the factor
$1\Bigl/ \lfloor {\bf{\cal A}}_3+ {\bf{\cal B}}_3\rfloor$  simply
becomes $1\Bigl/ \lfloor {\bf{\cal A}}_3- {\bf{\cal B}}_3\rfloor$
if it is carried over to the other side of the tensor product.
One further defines
$$
\Lambda(q^{{\bf{\cal A}}_3}) = q^{\demi{\bf{\cal A}}_3}
q^{-\demi{\bf{\cal B}}^3}
 \underbrace \otimes q^{\demi{\bf{\cal A}}_3}q^{\demi{\bf{\cal B}}_3}
$$
\beq
\Lambda(q^{{\bf{\cal B}}_3}) =  q^{\demi{\bf{\cal B}}_3}
q^{-\demi{\bf{\cal A}}_3}
 \underbrace  \otimes q^{\demi{\bf{\cal A}}_3}q^{\demi{\bf{\cal
B}}_3}
\label{coprodCJ}
\eeq
It now follows immediately that the algebra Eqs.
\ref{CAA}-\ref{ComAB}
is preserved by the above definition of the coproduct. Since the
matching
condition is automatically respected by the matrices under
consideration,
one can formally calculate without it,  taking into account only
Eq.\ref{match}.
On the other hand, the Casimir constraint Eq. \ref{relCAB}
enters explicitly.
We remark that the coproduct
takes the simpler form Eq. \ref{coprodAB} in terms of the matrices
$A^\lambda$ and $B^\lambda$, while their algebra is more
complicated than that of the ${\bf{\cal A}}_\pm,
{\bf{\cal B}}_\pm $. Note that the constraint  of equality between
the Casimir
eigenvalues  as well as the matching condition
are not compatible with the ordinary coproduct for
$U_q(sl(2)) \otimes U_q(sl(2))$. Once we have defined
a coproduct, the question naturally arises  whether we have
a full Hopf algebra. We will show, in section 7 that this is
not strictly the case, but that  an interesting Hopf like structure
of a novel
type arises which is a natural generalization of the usual one.

\subsection{Relation with the $\xi$ transformation laws}
\label{xitraforel}
Finally, we return to our beginning (section 3.1) to connect the
present
$U_q(sl(2)) \otimes U_q(sl(2))$ with the original $U_q(sl(2))$ that
appeared in the $\xi$ transformation laws. For this it is convenient
to
define matrices with four indices by formulae similar to
Eqs.\ref{newAB}. Consider  Eqs.\ref{Jtspsidef} and \ref{Jpmpsidef}.
It is convenient to let
$$
{\cal J}_\pm^\lambda]^{(J)}_{\rho'\varpi',\rho\varpi}=
\Jpmpsi J,\lambda,{{\varpi'-\rho'\over 2},},{{\varpi-\rho\over
2}},{\varpi'}
\delta_{\varpi',\varpi-2\lambda},\>
[q^{\pm {\cal J}_3\, \lambda}]^{(J)}_{\rho'\varpi',\rho\varpi}=
\Jtspsi J,\lambda,{{\varpi'-\rho'\over 2},},{{\varpi-\rho\over
2}},{\varpi'}
\delta_{\varpi',\varpi-2\lambda},
$$
\beq
[{\cal  X}_\pm^\lambda]^{(J)}_{\rho'\varpi',\rho\varpi}=
\Xpmpsi J,\lambda,{{\varpi'-\rho'\over 2},},{{\varpi-\rho\over
2}},{\varpi'}
\delta_{\varpi',\varpi-2\lambda},
\phantom{[q^{\pm  J_3 \, \lambda}]^{(J)}_{\rho'\varpi',\rho\varpi}=
\Jts J,\lambda,{{\varpi'-\rho'\over 2}},{{\varpi-\rho\over
2}},{\varpi'}
}.
\label{calmat}
\eeq
Then the relations Eqs.\ref{explma}, \ref{explma2}, \ref{Tpsidef},
\ref{defAB}, \ref{expAB}, \ref{defCAB} may be summarized by the
matrix equalities
\beqa
q^{{\cal J}_3\, \lambda}&=&
{2\lambda\over \lfloor {\cal A}_3-{\cal  B}_3\rfloor }
\left [  {\cal A}_{-2\lambda}+
{\cal B}_{-2\lambda}\right ]\nnn
{\cal X}_\pm^ \lambda&=&
{2\lambda q^{\pm 1/2}
\over \lfloor {\cal A}_3-{\cal  B}_3\rfloor } \left [
q^{\pm 2\lambda( {\cal A}_3-{\cal  B}_3)} {\cal A}_{-2\lambda}+
q^{\mp 2\lambda( {\cal A}_3-{\cal  B}_3)}
{\cal B}_{-2\lambda}\right ], \nnn
{\cal J}_\pm^ \lambda&=&
{\pm q^{\mp  1/2}\over q-q^{-1}} \left [
{\cal X}^{\lambda}_\pm-
q^{\pm 2\lambda( {\cal A}_3+{\cal  B}_3)\pm\demi)} q^{- {\cal J}_3\,
\lambda}
\right ].
\label{relations1}
\eeqa
These relations imply some identities between these matrices, i.e.
$$
q^{-\demi} {\cal X}_+^ \lambda+ q^{\demi} {\cal X}_-^ \lambda
=
(q^{\Omega_L}+q^{-\Omega_L}) q^{ J_3 \, \lambda}
$$
\beq
=(q-q^{-1})({\cal J}_-^\lambda-{\cal J}_+^ \lambda)+
(q^{\Omega_R}+q^{-\Omega_R}) q^{ -J_3 \, \lambda}
\label{calc5m}
\eeq
 For consistency we finally show  how  the algebra
of the matrices on the left hand sides, which follow  from
the definitions Eqs.\ref{Jtspsidef}, \ref{Jpmpsidef}, \ref{Xdef},
  may  be derived from
Eqs.\ref{CAA}--\ref{relCAB}. For this we will use the matrix
equivalent
of Eqs.\ref{Jtspsidef}, \ref{Jpmpsidef}. Let us define
\beqa
&[{ J}_\pm^\lambda]^{(J)}_{\rho'\varpi',\rho\varpi}=
\Jpm {},\lambda,{{\varpi'-\rho'\over 2},},{{\varpi-\rho\over
2}},{\varpi'}
\delta_{\varpi',\varpi-2\lambda},\quad
&[q^{\pm  J_3 \, \lambda}]^{(J)}_{\rho'\varpi',\rho\varpi}=
\Jts {},\lambda,{{\varpi'-\rho'\over 2},},{{\varpi-\rho\over
2}},{\varpi'}
\delta_{\varpi',\varpi-2\lambda},\nnn
&[ {\cal U}]^{(J)}_{\rho'\varpi',\rho\varpi}=
\UU J,{\varpi},{{\varpi'-\rho'\over 2},},{{\varpi-\rho\over
2}},{\varpi'}
\> \delta_{\varpi',\varpi},\quad
&[{\cal V}]^{(J)}_{\rho'\varpi',\rho\varpi}=
\VV J,{\varpi},{{\varpi'-\rho'\over 2},},{{\varpi-\rho\over
2}},{\varpi'}
\> \delta_{\varpi',\varpi}.
\label{matdef}
\eeqa
On the r.h.s. of the first line, we use the $U_q(sl(2))$ matrix
representation
that appeared in section 2.
It follows from the definition Eqs.\ref{xipsi} that
 the last two matrices are inverses of one another
\beq
\sum_{\rho_2, \varpi_2} [ {\cal
U}^\lambda]^{(J)}_{\rho_3\varpi_3,\rho_2\varpi_2}
[ {\cal V}^\lambda]^{(J)}_{\rho_2\varpi_2,\rho_1\varpi_1}=
\delta_{\varpi_3,\varpi_1} \delta_{\rho_3,\rho_1},
\label{inv}
\eeq
so that Eqs.\ref{Jtspsidef}, \ref{Jpmpsidef} take the matrix forms
\beq
q^{\pm {\cal J}_3\, \lambda}= {\cal U}
q^{\pm {J}_3\, \lambda} {\cal U}^{-1}, \qquad
{\cal J}_\pm^\lambda = {\cal U}
{J}_\pm^\lambda  {\cal U}^{-1}.
\label{simil}
\eeq
 So far we defined
our matrices with indices $\lambda=\pm 1/2$. The definitions given
immediately extend to arbitrary $m/2$ with $m$ integer. Then
 it is easy to see that the matrices ${J}_\pm^{m/2} $, and
$q^{\pm {J}_3 , {m/2}}$, $q^{\Omega_R}$  generate a loop extension
of $U_q( sl(2))$, that is
\beqa
\Bigl [ {J}_+^{m\over 2}, {J}_-^{n\over 2}  \Bigr]&=&
{q^{ {2J}_3,\, {m+n\over 2}}-q^{-{2J}_3,\, {m+n\over 2}}
\over q-q^{-1}},\quad
q^{ {2J}_3, \, {m+n\over 2}}=
q^{ {J}_3,\, {m\over 2}}q^{ {J}_3,\, {n\over 2}},\quad
q^{ {J}_3,\, {m\over 2}}q^{- {J}_3,\, -{m\over 2}} =1,\nnn
q^{ {J}_3,\, {m\over 2}}  J_\pm ^{n\over 2}&=&q^{\pm 1}
J_\pm^{n\over 2} q^{ {J}_3\, {m\over 2}},
\> q^{\Omega_R} J_\pm ^{m\over 2}=
 J_\pm ^{m\over 2} q^{\Omega_R-m},
\> q^{\Omega_R} q^{\pm J_3,\, {m\over 2}}
=   q^{\pm J_3,\, {m\over 2}}  q^{\Omega_R-m}.
\eeqa
Thus $ q^{\Omega_R}$ is the grading operator.
Moreover, if we introduce the matrix
\beq
[{ I}]_{\rho'\varpi',\rho\varpi}=
\delta_{{\varpi'-\rho'\over 2}, \, {\varpi'-\rho'\over 2}}
\delta_{\varpi', \varpi-1},
\label{Idef}
\eeq
we obviously have
\beqa
{J}_+^{m\over 2}&=&{J}_+^0 \> I^{m},
\quad q^{ {J}_3, \, {m\over 2}}= q^{ {J}_3, \,0 }\> I^{{m}},
\nnn
{J}_+^{m\over 2}\>  I&=& I \> {J}_+^{m\over 2} , \quad
q^{ {J}_3, \, {m\over 2}}\> I= I\>  q^{ {J}_3, \, {m\over 2}},
\label{relations3}
\eeqa
where ${J}_+^0$, and $q^{ {J}_3\,0 }$ are the usual matrices of
$U_q(sl(2))$. They commute with $q^{\Omega_R}$ and $I$, which satisfy
\beq
q^{\Omega_R} I=  I\> q^{\Omega_R+1}.
\label{relations4}
\eeq
Clearly, in view of the similarity
relation  Eq.\ref{simil}, if we let
\beq
{\cal I}= {\cal U}
I {\cal U}^{-1},
\label{calIdef}
\eeq
the matrices ${\cal J}_\pm^{m\over 2}$
 $q^{ {\cal J}_3, \,{m\over 2} }$  $ q^{\Omega_R}$
$\cal I$ also
satisfy the algebra\footnote{ $q^{\Omega_R}$ is invariant under
the transformation.} just displayed.   Next we
connect its restriction for  $\lambda=\pm 1/2$ to the
$U_q(sl(2))\otimes U_q(sl(2))$ algebra  Eqs.\ref{CAA}--\ref{relCAB}
by
making use of Eqs.\ref{relations1}.
Note that  the first two equalities of \ref{relations1}
are simple while the last one involves
$q^{- {\cal J}_3\, - \lambda}$ which, being   the inverse of
$q^{ {\cal J}_3\,  \lambda}$, is to be computed by taking the inverse
of
both sides of the first equation.  Consistently with that, and
by a mechanism which resembles the use of Eq.\ref{D2} to eliminate
${\cal O}\left [q^{-J_3}\right]$ (recalled in section 2), we will see
that
if we consider the algebra of  $ q^{ {J}_3\,\mu }$ and
${\cal X}_\pm^ \lambda$
(instead of $ q^{ {J}_3\,\mu }$  and
 ${\cal J}_\pm^ \lambda$),  the complicated operator
$q^{ {-J}_3\,\lambda }$
will disappear from the algebra.
For this, one first deduces the following relations
satisfied by   the $\chi$ matrices
\beqa
q^{J_3, \, \lambda} \chi_\pm ^{\mu} - q^{\pm 1} \chi_\pm ^{\mu}
q^{J_3, \, \lambda} &=&
\mp (q-q^{-1}) q^{\mp (2\lambda \Omega_R-\demi)} \delta_{\lambda+\nu,
\, 0}
\nnn
q^{J_3, \, \lambda} \chi_\pm ^{\mu} - q^{\pm 1} \chi_\pm ^{\lambda}
q^{J_3, \, \mu} &=&
 (q-q^{-1})
\left [q^{\mp (2\lambda \Omega_R-\demi)}-
q^{\pm (2\lambda \Omega_R+\demi)}\right] \delta_{\lambda+\nu, \, 0}
\label{q-chi}
\eeqa
$$
\Bigl [ \chi_+  ^{\mu}, \, \chi_- ^{\lambda} \Bigr ] =
-(q-q^{-1}) q^{J_3, \, \mu  }q^{J_3, \, \lambda}
$$
$$
+(q-q^{-1}) q^{- 2\lambda \Omega_R} (q^{-\demi} \chi_+  ^{\lambda} +
q^{\demi} \chi_-  ^{\lambda}) q^{-J_3, \, \lambda}
\delta_{\lambda+\nu, \, 0}
$$
$$
\chi_+  ^{\mu}\, \chi_- ^{\lambda}-
 \chi_+  ^{\lambda} \, \chi_- ^{\mu} =
-(q-q^{-1}) q^{J_3, \, \mu  }q^{J_3, \, \lambda}
$$
\beq
-[q^{ 2\lambda \Omega_R-1}-q^{- 2\lambda \Omega_R+1}]
 (q^{-\demi} \chi_+  ^{\lambda} +
q^{\demi} \chi_-  ^{\lambda}) q^{-J_3, \, \lambda}
\delta_{\lambda+\nu, \, 0}
\label{chi-chi}
\eeq
\beq
\Bigl [ \chi_+  ^{\mu}, \, \chi_+ ^{\lambda} \Bigr ]=
\Bigl [ \chi_-  ^{\mu}, \, \chi_- ^{\lambda} \Bigr ]=0
\label{chi-chi0}
\eeq
Using Eqs.\ref{calc5m}, Eqs.\ref{chi-chi} become
$$
\Bigl [ \chi_+  ^{\mu}, \, \chi_- ^{\lambda} \Bigr ] =
-(q-q^{-1}) q^{J_3, \, \mu  }q^{J_3, \, \lambda}
+(q-q^{-1})
[q^{-2\lambda\Omega_R}(q^{\Omega_L}+q^{- \Omega_L})]
\delta_{\lambda+\nu, \, 0}
$$
\beq
\chi_+  ^{\mu}\, \chi_- ^{\lambda}-
 \chi_- ^{\mu} \, \chi_+  ^{\lambda} =
-(q-q^{-1}) q^{J_3, \, \mu  }q^{J_3, \, \lambda}
-[q^{ 2\lambda \Omega_R-1}-q^{- 2\lambda \Omega_R+1}]
[q^{ \Omega_L}+q^{- \Omega_L}]
\delta_{\lambda+\nu, \, 0}.
\label{chi-chi1}
\eeq
It is now straightforward to verify that Eqs.\ref{q-chi},
\ref{chi-chi0},
\ref{chi-chi1}, as well as the relations
\beq
\left[ q^{J_3,\, \lambda},\, q^{J_3,\, \mu}\right]=0
\label{chi-chi2}
\eeq
follow from the $U_q(sl(2))\otimes U_q(sl(2))$ relations displayed by
 Eqs.\ref{CAA}--\ref{relCAB}
if one uses the first two relations of  Eqs.\ref{relations1}.
Moreover, Eqs. \ref{relations1} can be used to rewrite the $\cal A$
and $\cal B$ matrices as
$$
{\cal A}_{2\lambda}={1\over 2} \left \{\lfloor \Omega_L\rfloor
q^{{\cal J}_3,-\lambda} +
\lfloor \Omega_R\rfloor
q^{-{\cal J}_3,-\lambda} -2\lambda
({\cal J}_+^ {-\lambda}+ {\cal J}_-^ {-\lambda})
\right\}
$$
\beq
{\cal B}_{2\lambda}=
{1\over 2} \left \{\lfloor \Omega_L\rfloor
q^{{\cal J}_3,-\lambda} -
\lfloor \Omega_R\rfloor
q^{-{\cal J}_3,-\lambda} +2\lambda
({\cal J}_+^ {-\lambda}+ {\cal J}_-^ {-\lambda})
\right\}
\label{xiAB}
\eeq
Thus the six operators $q^{{\cal A}_3}$, $q^{{\cal B}_3}$, ${\cal
A}_\pm$,
 ${\cal B}_\pm$ are functions of five  operators, e.g. $\Omega_L$,
$\Omega_R$, $q^{{\cal J}_3,\pm\demi}$,
 ${\cal J}_+^ {\demi}+ {\cal J}_-^ {\demi}$, ${\cal I}$ (the latter
is defined
by Eq.\ref{calIdef}). So there
must be one constraint relating them, which is precisely the equality
of the Casimir operators discussed above.
\subsection{Internal invariance}
The aim of this subsection is to show that the matrices $A^\lambda$,
$B^\lambda$, $q^{\Omega_L}$,
$q^{\Omega_R}$ introduced above, which describe
the operatorial actions Eqs.\ref{2TactAB}, \ref{shift},
\ref{Ttsopdef},
give rise to symmetries of the operator algebra of the Bloch waves.
The reasoning we apply is closely parallel to the one for the
covariant basis,
which was laid out in ref. \cite{CGS1}. Let us therefore recall
briefly
the situation for the covariant fields. Consider Eqs.\ref{act+},
 \ref{act3+}. For each of the first two operator actions, two
matricial
actions, $\left[q^{-J_3}\right]_{NM}$, $\left[J_+\right]_{NM}$ and
$\left[q^{-J_3}\right]_{NM}$, $\left[J_-\right]_{NM}$ appear.
They are multiplied by different operators, and thus lead
to independent symmetries
of the operator algebra. This is easily seen upon combining
the operatorial actions with the commutativity of
fusion and braiding, and the Yang-Baxter equation,
two special cases of the general Moore-Seiberg consistency conditions
\cite{MS}. Since $\left[q^{-J_3}\right]_{NM}$ appears twice, we have
three symmetries $\left[J_\pm\right]_{NM}$, $\left[ J_3\right]_{NM}$
altogether, in correspondence with the number of generators.
More precisely,
the following matricial, or `internal' transformations:
\beq
\xi_M^{(J)}(\sigma) \to
\sum _N \xi_N^{(J)}(\sigma) \left [J^a\right]_{NM}.
\label{qact}
\eeq
\beq
\xi^{(J_1)}_{M_1}(\sigma_1) \xi^{(J_2)}_{M_2}(\sigma_2) \to
\sum _{N_1,N_2,b,c}
\xi^{(J_1)}_{N_1}(\sigma_1) \xi^{(J_2)}_{N_2}(\sigma_2)
\Lambda_{bc}^a \left[J^b\right]_{N_1 M_1}
\left[J^c\right]_{N_2 M_2},
\label{xicoprod2}
\eeq
preserve the fusion and the braiding of the $\xi$ fields.
The fusion of the $\xi$ fields is given essentially
in terms of $q-3j$ symbols,
\beq
\xi ^{(J_1)}_{M_1}(\sigma_1)\,\xi^{(J_2)}_{M_2}(\sigma_2) =
 \sum _{J_{12}= \vert J_1 - J_2 \vert} ^{J_1+J_2}
g _{J_1J_2}^{J_{12}} (J_1,M_1;J_2,M_2\vert J_{12})
 (\xi ^{(J_{12})} _{M_1+M_2}(\sigma_2) + \hbox{desc.}),
\label{fusxi}
\eeq
where $g^{J_{12}}_{J_1J_2}$ are coupling constants (reduced matrix
elements) and
desc. denotes Virasoro descendants. The fusion and braiding
properties
of the latter are the same as those of the primaries \cite{CGR2}, so
we will
not consider them explicitly in the following.
The quantum group invariance of the fusion is equivalent to the
standard
recursion relations for the $q-$ Clebsch-Gordan coefficients,
$$
\sum_{N_1+N_2=N_{12}}  (J_1,N_1;J_2,N_2\vert J_{12})
\Lambda_{de}^b \left[J^d\right]
_{N_1M_1}
\left[J^e\right]_{N_2M_2}
=
$$
\beq
(J_1,M_1;J_2,M_2\vert J_{12}) \left[J^b\right]_{N_{12}M_{12}}.
\label{rec3j}
\eeq
Similarly, as the braiding of two $\xi$ fields is given by the
universal
$R$-matrix of $U_q(sl(2))$, the invariance of the braiding is
tantamount
to the defining relation between co-product and $R$-matrix, viz.
\beq
(J_1,J_2)_{N_1\, N_2}^{P_2\, P_1} \Lambda_{de}^b
\left[J^d\right]_{N_1M_1} \left
[J^e\right]_{N_2M_2}=
\Lambda_{de}^b \left[J^d\right]_{P_2N_2} \left[J^e\right]_{P_1N_1}
(J_1,J_2)_{M_1\, M_2}^{N_2\, N_1}.
\label{defunvR}
\eeq
Eq. \ref{defunvR} arises in our operator formalism as a direct
consequence of the Yang-Baxter
equation.

Let us now examine the case of the $\psi$ fields and look for
an internal type invariance of the fusion and the braiding.
We start from the fusion relation for the $\psi$ fields
($m_{12}=m_1+m_2$),
\beq
{\psi}^{(J_1)}_{m_1}(\sigma_1)
{\psi}^{(J_2)}_{m_2}(\sigma_2)
=
\sum _{J_{12}= -m_{12}} ^{J_1+J_2}
g_{J_1J_2}^{J_{12}}\ N\left|
^{J_1}_{m_1}\,^{J_2}_{m_2}\,^{J_{12}}_{m_1 +m_2},{\varpi} \right| \>
({\psi}^{(J_{12})}_{m_{12}}(\sigma_2) + \hbox{desc.}),
\label{psifus}
\eeq
where $g^{J_{12}}_{J_1 J_2}$
are the same coupling constants as in Eq. \ref{fusxi},
and the fusion coefficient $N$ is given in terms of a $q-6j$
symbol\footnote{
The condition $m_{12}=m_1+m_2$ is not implied by the property of the
6j
symbols and should be added by hand}
,
\beq
N\left| ^{J_1}_{m_1}\,^{J_2}_{m_2}\,^{J_{12}}_{m_{12}},{\varpi}
\right| \>=
 {\gtil^x_{J_{12} x+m_{12}} \over \gtil^x_{J_1 x+m_1}
\gtil ^{x+m_1}_{J_2 x+m_{12}}}
\left\{ ^{ J_1}_{x+m_{12}}\,^{ J_2}_{x}
\right. \left |^{J_{12}}_{x+m_1}\right\}.
\label{defN}
\eeq
Here we have introduced the abbreviations $m_{12}=m_1+m_2,
n_{12}=n_1+n_2$
and, as in previous references,
\beq
x:=\demi(\varpi-1-\pi/h).
\eeq
The constants  $\gtil$ can be absorbed into the normalization of the
$\psi^{(J)}
_m$ and will not play any significant role.
We recall that the $6j$ coefficient
 with a vertical bar used here and in previous references is related
to the
tetrahedron-symmetric $q-6j$ symbol by
\beq
\left\{  ^a _d \, ^b _c \, ^e _f \right\}=
(\lfloor2e+1\rfloor \lfloor 2f+1\rfloor)^{-\demi}(-1)^{a+b-c-d-2e}
\left\{^a_d \, ^b_c \right.\left |^e_f \right\}.
\eeq
Let us introduce a $3j$ symbol with two magnetic quantum numbers
by\footnote{
The conservation of $m$ imposed earlier now  follows from a
combination of the
three delta functions.}
$$
(J_1\rho_1\varpi_1;J_2\rho_2\varpi_2 |J_{12}\rho_{12}\varpi_{12}):=
 \delta_{\rho_{12},\rho_1}\delta_{\varpi_{12},\varpi_2}
\delta_{\varpi_1,\rho_2}\times
$$
\beq
N\left \vert ^{J_1} _{{\varpi_1-\rho_1 \over 2}}
 \, ^{J_2}_{{\varpi_2-\rho_2 \over 2}}
 \, ^{J_{12}} _{{\varpi_{12}-\rho_{12}\over 2}} ,
 \, \rho_1 \right \vert
\label{new3j}
\eeq
Then Eq.\ref{psifus} can be rewritten as
$$
\psi^{(J_1)}_{\rho_1,\varpi_1}\psi^{(J_2)}_{\rho_2,\varpi_2}
\delta_{\varpi_1,\rho_2}
=
\sum_{J_{12}=-m_{12}}^{J_1+J_2}
g^{J_{12}}_{J_1 J_2} (J_1,\rho_1,\varpi_1;J_2,\rho_2,
\varpi_2 |J_{12},\rho_{12},\varpi_{12})
 \times
$$
\beq
(\psi^{(J_{12})}_{\rho_{12},\varpi_{12}} +\hbox{desc.}).
\label{newfus}
\eeq
Let us now braid both sides of Eq.\ref{newfus} with
$\Top{+}, \lambda, {\sigma_+} $, and then fuse again the result
on the left hand side to $\psi^{(J_{12})}_{\rho_{12},\varpi_{12}}$.
 Using Eq.\ref{2TactAB}, we get
$$
\sum_{\rho_1'\varpi'_1,\rho'_2\varpi'_2}
(J_1,\rho'_1,\varpi'_1;J_2,\rho'_2,\varpi'_2 \vert
J_{12},\rho'_{12},\varpi'_{12} )
\{ \Top{+}, \lambda, {\sigma_+}
(\An {J_1},{\lambda},{\rho'_1},{\varpi'_1},{\rho_1},{\varpi_1}
 \An {J_2},{\lambda},{\rho'_2},{\varpi'_2},{\rho_2},{\varpi_2}
$$
$$
+\Bn {J_1},{-\lambda},{\rho'_1},{\varpi'_1},{\rho_1},{\varpi_1}
 \Bn {J_2},{\lambda},{\rho'_2},{\varpi'_2},{\rho_2},{\varpi_2}  )
+ \Top{-}, \lambda, {\sigma_+}
(\An {J_1},{\lambda},{\rho'_1},{\varpi'_1},{\rho_1},{\varpi_1}
 \Bn {J_2},{-\lambda},{\rho'_2},{\varpi'_2},{\rho_2},{\varpi_2}
$$
$$
+ \Bn {J_1},{-\lambda},{\rho'_1},{\varpi'_1},{\rho_1},{\varpi_1}
  \An {J_2},{-\lambda},{\rho'_2},{\varpi'_2},{\rho_2},{\varpi_2}  )
\} =
$$
$$
(J_1,\rho_1,\varpi_1;J_2,\rho_2,\varpi_2 \vert
J_{12},\rho_{12},\varpi_{12} ) \{
\Top{+}, \lambda, {\sigma_+}
\An
{J_{12}},{\lambda},{\rho'_{12}},{\varpi'_{12}},
{\rho_{12}},{\varpi_{12}}
$$
\beq
+
\Top{-}, \lambda, {\sigma_+}
\Bn
{J_{12}},{-\lambda},{\rho'_{12}},{\varpi'_{12}},
{\rho_{12}},{\varpi_{12}}
\}
\eeq
Exactly as for the case of the covariant fields, we now compare the
coefficents
of like operators and obtain
$$
\sum_{\rho_1'\varpi'_1,\rho'_2\varpi'_2}
(J_1,\rho'_1,\varpi'_1;J_2,\rho'_2,\varpi'_2 \vert
J_{12},\rho'_{12},\varpi'_{12} )
\Lambda(A^\lambda)_{\rho'_1\varpi'_1,\rho'_2\varpi'_2;
\rho_1\varpi_1,\rho_2\varpi_2}=
$$
\beq
\An
{J_{12}},{\lambda},{\rho'_{12}},{\varpi'_{12}},
{\rho_{12}},{\varpi_{12}}
(J_1,\rho_1,\varpi_1;J_2,\rho_2,\varpi_2 \vert
J_{12},\rho_{12},\varpi_{12} )
\label{fusinv}
\eeq
where Eq.\ref{copA} has been used, and similarly with $A^\lambda$
replaced
by $B^\lambda$. Recalling that the generalized $3j$ symbols above are
given in terms of $q-6j$ symbols,
Eq. \ref{fusinv} demonstrates that for the $\psi$ basis, the $6j$
symbols acquire an
interpretation as Clebsch-Gordan coefficients for the new quantum
group structure.
We can use Eq. \ref{fusinv} to determine the new $3j$  symbols by
recursion, in analogy to
the standard case, except that here we need to know, for example, the
coefficient
$(J_1,\varpi_1-2J_1,\varpi_1;J_2,\varpi_2-2J_2,
\varpi_2|J_{12},\rho_{12},\varpi_{12})$  for all $\varpi_2$ as a  
starting point. Eq.\ref{fusinv} arises
from the commutativity of fusion and braiding - one of the
Moore-Seiberg
consistency conditions - exactly as in the covariant basis.
We have thus seen that
the fusion is invariant under the internal transformations
\beq
\psi^{(J)}_{\rho,\varpi}  \to
\sum_{\rho',\varpi'}  \psi ^{(J)}_{\rho',\varpi'}
[A^\lambda]^{(J)}_{\rho'\varpi',\rho\varpi}
\label{defAV}
\eeq
and
\beq
\psi^{(J)}_{\rho,\varpi}  \to
\sum_{\rho',\varpi'}  \psi ^{(J)}_{\rho',\varpi'}
[B^\lambda]^{(J)}_{\rho'\varpi',\rho\varpi},
\label{defBV}
\eeq
together with the co-product defined by Eqs.\ref{3TactAB}.
Concerning the zero modes, we have
two additional internal invariances which leave the fusion invariant:
\beq
\psi ^{(J)}_{\rho,\varpi}\to \psi ^{(J)}_{\rho,\varpi} q^\rho, \quad
\psi ^{(J)}_{\rho,\varpi}\to \psi ^{(J)}_{\rho,\varpi} q^\varpi,
\label{defOmegaV}
\eeq
corresponding to the matrices $q^{\Omega_L}$ and
$q^{\Omega_R}$ respectively (see Eq.\ref{Omegdef}).
The action on $\psi \otimes \psi$  is given
by their co-product displayed on Eq.\ref{coprodCJ}.
Including $\Omega _L$ and $ \Omega _R$ gives six generators
for the internal
invariance. The Casimir constraint Eq.\ref{relCAB} tells us that only
five of them are independent.

A similar discussion applies to the braiding of the $\psi$ fields.
The general braiding relation reads \cite{CGR1}
$$
\psi^{(J_1)}_{m_1}(\sigma_1)\psi^{(J_2)}_{m_2}(\sigma_2)=
{\cal
S}(J_1,J_2;\varpi)^{m'_2m'_1}_{m_1m_2}\psi^{(J_2)}_{m'_2}(\sigma_2)
\psi^{(J_1)}_{m'_1}(\sigma_1),
$$
with
$$
{\cal S}(J_1,J_2;\varpi)^{m'_2m'_1}_{m_1m_2}=\sum_{m'_1,m'_2}
q^{\mp(2m_1m_2+
m_2^2-{m'_2}^2+\varpi(m_2-m'_2))}\delta_{{m'_1}+{m'_2},m_1+m_2}
{\gtil^{x+m'_2}_{J_1 x+m'_{12}}\gtil^x_{J_2,x+{m'_2}}\over
\gtil^{x+m_1}_{J_2,x+m_{12}}\gtil^x_{J_1,x+m_1}}
$$
\beq
\times\left\{ ^{J_1}_{J_2} \, ^{x+m_{12}}_x \right. \left |
^{x+m'_2}_{x+m_1}\right\}
\label{defS}
\eeq
and the upper sign  is to be taken when $\sigma_1>\sigma_2$. An
explicit
$\delta$ coefficient has been written for the conservation of $m$,
because
it is not automatically implied by the properties of the $6j$ symbol.
We note
that the braiding of descendants of the $\psi$ fields is given by the
same formula \cite{CGR1}. Again, we can rewrite this formula in terms
of
double index symbols:
\beq
\psi^{(J_1)}_{\rho_1 \varpi_1} \psi^{(J_2)}_{\varpi_1 \varpi_2}
=\sum_{ \varpi'_1 \rho'_2 \varpi'_2}
{\cal S}(J_1,J_2)_{\rho_1\varpi_1,\varpi_1 \varpi_2}^
{\rho'_2\varpi'_2,\varpi'_2\varpi'_1} \psi^{(J_2)}_{\rho'_2
\varpi'_2}
\psi^{(J_1)}_{\varpi'_2 \varpi'_1},
\label{excmat}
\eeq
The braiding matrix on the right hand side is given by
\beq
{\cal S}(J_1,J_2)_{\rho_1\varpi_1,
\rho_2\varpi_2}^{\rho'_2\varpi'_2,\rho'_1\varpi'_1}:=
{\cal S}(J_1,J_2,\varpi_2-2n_{12})_{m_1 m_2}^{m'_2 m'_1}
\delta_{\varpi_1,\rho_2}
\delta_{\varpi'_2,\rho'_1}
\delta_{\rho_1,\rho'_2}
\delta_{\varpi_2,\varpi'_1}
\label{newR}
\eeq
with
$$
m_1={\varpi_1-\rho_1\over 2}, m_2={\varpi_2-\rho_2\over 2},
m'_1={\varpi'_1-\rho'_1 \over 2}, m'_2={\varpi'_2-\rho'_2 \over 2}.
$$
The Kronecker symbols
$\delta_{\varpi_1,\rho_2}\delta_{\varpi'_2,\rho'_1}$
represent the matching conditions, while $\delta_{\rho_1,\rho'_2}
\delta_{\varpi_2,\varpi'_1}$ incorporate the conservation of $m$.
 The internal invariances of the braiding are now
generated by comparing the action of
$\Top{+}, \lambda, {\sigma_+} $ on a product of two $\psi$ fields
before and after the braiding of the latter. One obtains, by a
similar
argument as for the case of fusion,
$$
\sum_{\rho_3 \varpi_3, \rho_4 \varpi_4}
{\cal S}(J_1,J_2)^{\rho_6\varpi_6,\rho_5\varpi_5}_
{\rho_3\varpi_3,\rho_4\varpi_4}
\Lambda(A^\lambda)_{\rho_3\varpi_3,\rho_4\varpi_4;
\rho_1\varpi_1,\rho_2\varpi_2}
$$
\beq
=
\sum_{\rho_3 \varpi_3, \rho_4 \varpi_4}
{\cal S}(J_1,J_2)^{\rho_4\varpi_4,\rho_3\varpi_3}
_{\rho_1\varpi_1,\rho_2\varpi_2}
\bar\Lambda(A^\lambda)_{\rho_6\varpi_6,\rho_5\varpi_5;
\rho_4\varpi_4,\rho_3\varpi_3}
\label{newYB}
\eeq
where $\bar\Lambda$ is the co-product
with factors $1$ and $2$ exchanged, and similarly
for the case of $B^\lambda$. Of course, the commutativity of the two
orders
of the braiding is again one of the Moore-Seiberg conditions.
Eq. \ref{newYB} relates the co-product to the corresponding
$R$ matrix, again just as in the standard case,
so that Eq.\ref{newR} defines a kind of
`universal R matrix' associated with our quantum group structure.

Returning briefly to the $\xi$  fields, we remark that the action of
$U_q(sl(2)) \otimes U_q(sl(2))$ on them may be defined by using
Eqs.\ref{xiAB}.
It reduces to the  action recalled earlier by the matrices $J_\pm$,
$q^{J_3}$ (see Eqs.\ref{qact}, \ref{xicoprod2}), and the trivial
left and right multiplications by $q^\varpi$.
\section{Application to state/operator classifications}
Since our internal symmetry group is directly connected with
the Liouville zero mode, it should be a good tool to classify the
spectrum of primary fields and associated Verma modules.
We will discuss some aspects of this here, without going into
details.
\subsection{General aspects of the representations}
We will make use of the expressions Eqs.\ref{expAB}, \ref{defCAB}
for the generators. It will be convenient to simplify
notation by letting  $\mu=(\varpi+\rho)/2$, $\nu=(\varpi-\rho)/2$, so
that
\beqa
\left [ {\bf{\cal A}}_\pm\right ]_{\mu', \nu';\, \mu, \nu}&=&
\pm \lfloor \mu \pm (J+1)\rfloor
\delta_{\mu',\, \mu \pm 1} \delta_{\nu',\,\nu} \nnn
\left [ {\bf{\cal B}}_\pm\right ]_{\mu', \nu';\, \mu, \nu}&=&
\pm q^{\mp \demi}  \lfloor J\pm \nu \rfloor
\delta_{\mu',\, \mu } \delta_{\nu',\,\nu\pm 1}
\label{ABexp}
\eeqa
Note that, since the matrix elements are proportional to
q deformed numbers, one has
\beqa
\left [ {\bf{\cal A}}_\pm\right ]_{\mu', \nu';\, \mu, \nu}&=&
(-1)^\alpha
\left [ {\bf{\cal A}}_\pm\right ]_
{\mu'+\alpha\pish , \nu';\, \mu+\alpha\pish , \nu}
\nnn
\left [ {\bf{\cal B}}_\pm\right ]_{\mu', \nu';\, \mu, \nu}&=&
(-1)^\alpha
\left [ {\bf{\cal B}}_\pm\right ]_
{\mu', \nu'+\alpha\pish ;\, \mu, \nu+\alpha\pish},
\label{shiftcal}
\eeqa
where $\alpha$ is an arbitrary integer. This
reflects the existence of
 another ``dual'' quantum group with parameter $\hhat=\pi^2/h$
which commutes with the present one up to a sign. We will not
consider this group here, since we only include one screening
charge, but the equations just written are important for
the coming discussion where a shift of this type will be needed. To
shorten the
discussion we will only deal with positive half-integer spins $J$.
The generalization
is straightforward.
We will first discuss the case of generic $h$. Then it is
trivial to verify that
\beqa
\left [ {\bf{\cal A}}_-\right ]_{J, \nu;\, J+1, \nu}&=&
\left [ {\bf{\cal A}}_+\right ]_{-J, \nu;\, -(J+1), \nu}=0
\nnn
\left [ {\bf{\cal B}}_-\right ]_{\mu, -J-1;\, \mu, -J}&=&
\left [ {\bf{\cal B}}_+\right ]_{\mu, J+1;\, \mu, J}=0.
\label{maxw}
\eeqa
For generic $h$, these relations give us,  up to the shift
Eqs.\ref{shiftcal},
all the highest/lowest weight states.
 Consider first
the ${\bf{\cal A}}$ algebra, ignoring the $\nu$ index since these
generators do
no act upon it. The states with $\mu=(J+1)$, and  $\mu=-(J+1)$ are
lowest weight and
highest weight  states, respectively. For $J>0$, we get two disjoint
semi-infinite
representations with $\mu=J+1+n$ and $\mu=-(J+1+n)$, $n$ non-negative
integer, respectively. Next concerning the
${\cal B}$ algebra, the states with $\nu=J$ and  $\nu=-J$ are lowest
weight and
highest weight  states, respectively. For positive $J$, we get a
finite dimensional representation
if $2J$ is integer. Let us now show that these simple facts allow us
to recover the three
cases which were discussed earlier\cite{GR1}. There it was shown that
the type of operator
algebras is specified by the number---going from  one to  three---of
triangular
inequalities satisfied by each vertex.  These  cases
 were called TI1, TI2, TI3 in ref.\cite{GR1}. To specify the spins
associated with
$\varpi$ and $\rho$, one lets $\varpi=\varpi_0+2J_3$,
$\rho=\varpi_0+2J_2$, where
$\varpi_0=1+\pish$ is such that the corresponding eigenvalue of $L_0$
 vanishes.
In the TI3 case, one
has three triangular inequalities between $J$, $J_2$, $J_3$, so that
the latter  are all
half integers. This gives
$$
J+J_3-J_2\equiv J+(\varpi-\varpi_0)/2-(\rho-\varpi_0)/2=J+\nu \in
{\cal Z}_+
$$
$$
J-J_3+J_2\equiv J-(\varpi-\varpi_0)/2+(\rho-\varpi_0)/2=J-\nu \in
{\cal Z}_+
$$
\beq
-J+J_3+J_2\equiv
-J+(\varpi-\varpi_0)/2+(\rho-\varpi_0)/2=\mu-J-1-\pish \in {\cal
Z}_+.
\label{TI3}
\eeq
The first two inequalities give back  the range of the   finite
dimensional
 ${\bf{\cal B}}$ representation. For the matrices ${\bf{\cal A}}$ ,
we see that the lowest value
coincides with the lowest weight found above up to a shift of
$\pish$, which may be incorporated
easily by using Eq.\ref{shiftcal}.
Thus the last inequality corresponds to the  semi-infinite
representation with lowest weight which we found earlier. The other
semi-infinite ${\bf{\cal A}}$ representation is easily seen to
correspond to
negative spins using the formulae just recalled, and we leave it out
for the
present time. For the case TI2, one only imposes
$$
J+J_3-J_2\equiv J+(\varpi-\varpi_0)/2-(\rho-\varpi_0)/2=J+\nu \in
{\cal Z}_+
$$
$$
J-J_3+J_2\equiv J-(\varpi-\varpi_0)/2+(\rho-\varpi_0)/2=J-\nu \in
{\cal Z}_+
$$
Thus the  ${\bf{\cal B}}$ representation is still finite dimensional,
while the
${\bf{\cal A}}$ representation has no lower or upper bound. This is
consistent with the above
discussion because for the latter the lowest weight state is never
reached
since $J_2+J_3$ is not half integer.
For the TI1 case, that is for
$$
J+J_3-J_2\equiv J+(\varpi-\varpi_0)/2-(\rho-\varpi_0)/2=J+\nu \in
{\cal Z}_+
$$
the  ${\bf{\cal B}}$ representation is semi-infinite. Again this is
in agreement with the above.

Next, it is interesting to consider the particular case $J=0$. There
are two possibilities,
First, if we use the formulae with
$2J$ integer, and let $J=0$, the ${\bf{\cal B}}$ representation has
dimension one, and is
restricted to $\nu=0$.
 For the
${\bf{\cal A}}$ representation, we get the range $\mu\geq \varpi_0$.
Thus the spin zero
representation
is non trivial. As we will see, in section 7,  this explains why our
coproduct does not
possess a counit in the usual sense.
Another type of spin zero representation is obtained by considering
the TI1 case, where $J$
may take continuous values, and letting $J\to 0$. Then the ${\bf{\cal
B}}$ representation is
semi-infinite. This corresponds to the case of the powers of the
screening operator which will be
studied in the forthcoming subsection.

Finally let us briefly turn to the case where $q$ is a root of unity.
 For a rational theory, with $C=1-6(p-p')^2/pp'$, the spectrum of
Virasoro weights is
given by
$$
\Delta_{r,t}={(p'r-pt)^2-(p-p')^2\over 4pp'}.
$$
The correspondence with the present formalism is such that we have
$h=-p'\pi/p$, and
$r=2J+1$, $t=2\Jhat+1$, where $\Jhat$ specifies the representation of
the
dual quantum group with parameter $\hhat=-p\pi/p'$. As shown by BPZ,
the set of primary
fields with $1\leq r\leq p-1$, $1\leq t\leq p'$, $r$, $t$ integers
form a closed OPA,
if one
indentifies the operators with
quantum numbers $(r,t)$ and $(p-r, \, p'-t)$.
 Let us show that our representation theory correctly
gives the corresponding truncation of the spectrum of zero modes,
assuming that we consider
our $\psi$ fields with $0\leq 2J\leq p-2$ according to the limits
just recalled.
Since we do not include
the second screening charge, we  only discuss the case
$t=1$, i.e. $\Jhat=0$.  Consider the  ${\bf{\cal A}}$ representation,
with lowest weight
vector $\mu =\varpi_0+J$. Making use of the explicit expression
Eqs.\ref{ABexp},
one sees that
$$\left [ \left({\bf{\cal A}}_+\right)^{p-1-2J}\right ]^{(J)}_
{\varpi_0+p-1-J, \, \nu;\, \varpi_0+J,\, \nu}=0.
$$
Thus the range of $\mu$ is $\varpi_0 +J\leq \mu \leq \varpi_0+p-1-J$.
Concerning the ${\bf{\cal B}}$
representation, it is easily sees that the range is still $-J\leq \nu
\leq J$. Returning to
$\rho$ and $\varpi$, one verifies that one has $\varpi_0\leq \varpi
\leq \varpi_0+p-2$,
 $\varpi_0\leq \rho \leq \varpi_0+p-2$. Letting again
$\varpi=\varpi_0+2J_3$,
$\rho=\varpi_0+2J_2$, one sees that $J_2$ and $J_3$ vary over the
same range as $J$,
which is what we wanted to prove. There are of course many more
points to discuss, such as
the interpretation of the other representations, but we leave them
for further study.
\subsection{The internal symmetries of the Coulomb gas operators}
As another application  we derive the
transformation laws of the Coulomb gas  operators:  the
B\"acklund free field and  the screening operators. This will provide
a concrete
realization of the spin zero representation mentioned above. Since we
have to
change the operator normalizations we return to the earlier
formulation using the
notation $\psi_m^{(J)}$. It is easy to derive
the explicit form of the action of ${\cal O}[T_-]_{\sigma_+}$ from
 Eqs.\ref{defAB}, \ref{expAB}. Next,
we need to go from the $\psi$ fields to the $U$ fields introduced in
ref.\cite{CGR1} whose normalization is well suited for discussing
Coulomb gas operators. The correspondence is
given by
\beq
\psi^{(J)}_m(\sigma)={1\over \beta_m^{(J)} \gamma_m^{(J)}}
U^{(J)}_m(\sigma)
\label{psi-U}
\eeq
where
$$
\beta_m^{(J)}=e^{ih(J+m)(\varpi-J+m)}
$$
$$
\gamma_m^{(J)}=\mu^{J+m}{\rho(\varpi)\over \rho(\varpi+2m)}
\prod_1^{J+m} \lfloor \varpi+r\rfloor \lfloor \varpi+2m-r\rfloor
$$
$$
\rho=\sqrt{ \Gamma(\varpi h/ \pi) \Gamma(\varpi+1)
\Gamma_q(\varpi+1)},
\quad \mu=-{\pi^2\over h\sin h}
$$
$\Gamma_q$ is the q deformed gamma function.
This form is valid for arbitrary $J$
provided $J+m$ is integer\cite{GS3}.
We get rid of $\rho$ by transforming all the fields
including the
${\cal O}[T_\pm]_{\sigma_+}$ generators,
and forget about it. The action of the latter is
straightforwardly deduced from the formulae just given:
\beqa
{\cal O}[T_-]_{\sigma_+} U^{(J)}_m(\sigma)&=&
U^{(J)}_m(\sigma) e^{-ih(J+m)}{\lfloor \varpi-2m\rfloor \over \lfloor
\varpi\rfloor}
  {\lfloor \varpi-J-m-1\rfloor \over \lfloor \varpi-1\rfloor} {\cal
O}[T_-]_{\sigma_+}\nnn
&+& \mu U_{m-1}^{(J)}(\sigma) e^{ih(\varpi-J-m-\demi)}
\lfloor \varpi-2m+2\rfloor \lfloor J+m\rfloor {\cal
O}[T_+]_{\sigma_+},
\nnn
{\cal O}[T_+]_{\sigma_+} U^{(J)}_m(\sigma)&=&
U^{(J)}_m(\sigma)e^{ih(J+m)}
  {\lfloor \varpi+J-m+1\rfloor \over \lfloor \varpi-2m+1\rfloor}
{\cal O}[T_+]_{\sigma_+}\nnn
&-&\mu^{-1} U^{(J)}_{m+1}(\sigma) e^{-ih(\varpi-J-m-{3\over 2})}
{\lfloor J-m\rfloor \over \lfloor \varpi\rfloor \lfloor
\varpi-1\rfloor
\lfloor \varpi-2m-1\rfloor}
{\cal O}[T_-]_{\sigma_+}.\nnn
&&
\label{Utr}
\eeqa
Next, letting $J=0$ gives the transformation properties of the
screening operators
\beqa
{\cal O}[T_-]_{\sigma_+} S^m(\sigma)&=&
S^m(\sigma) e^{-ihm}{\lfloor \varpi-2m\rfloor \over \lfloor
\varpi\rfloor}
  {\lfloor \varpi-m-1\rfloor \over \lfloor \varpi-1\rfloor} {\cal
O}[T_-]_{\sigma_+}\nnn
&+& \mu S^{m-1}(\sigma) e^{ih(\varpi-m-\demi)} \lfloor
\varpi-2m+2\rfloor \lfloor m\rfloor {\cal O}[T_+]_{\sigma_+},
\nnn
{\cal O}[T_+]_{\sigma_+} S^m(\sigma)&=&
S^m(\sigma)e^{ihm}
  {\lfloor \varpi-m+1\rfloor \over \lfloor \varpi-2m+1\rfloor} {\cal
O}[T_+]_{\sigma_+}\nnn
&+& \mu^{-1} S^{m+1}(\sigma) e^{-ih(\varpi-m-{3\over 2})}
{\lfloor m\rfloor \over \lfloor \varpi\rfloor \lfloor \varpi-1\rfloor
\lfloor \varpi-2m-1\rfloor}
{\cal O}[T_-]_{\sigma_+}.\nnn
&&
\label{Str}
\eeqa
As a check, one may verify that these formulae are consistent with
the simple  fusion algebra
$S^m S^p\sim S^{m+p}$, which shows that these operators are indeed
powers of the screening operator
$S$. They provide a concrete realization of the spin zero
representation of our
internal symmetry group, where the ${\bf {\cal B}}$ representation is
semi-infinite.

Finally, the transformation laws of the B\"acklund free field are
obtained by expanding in J  for
fixed screening number $n=J+m$, according to $U^{(J)}_m\sim
(1-\alpha_-J\vartheta)S^{n}$. Before
expanding, it is useful to remark that most of the  explicit
dependence upon J for fixed $J+m$ may be
removed by moving some factors to the left, and rewriting
Eq.\ref{Utr} as
according to
\beqa
{\cal O}[T_-]_{\sigma_+} U^{(J)}_m(\sigma)&=&\lfloor \varpi\rfloor
U^{(J)}_m(\sigma) e^{-ih(J+m)}{1 \over \lfloor \varpi\rfloor}
  {\lfloor \varpi-J-m-1\rfloor \over \lfloor \varpi-1\rfloor} {\cal
O}[T_-]_{\sigma_+}\nnn
&+&\mu \lfloor \varpi\rfloor  U_{m-1}^{(J)}(\sigma)
e^{ih(\varpi-J-m-\demi)}
\lfloor J+m\rfloor {\cal O}[T_+]_{\sigma_+},
\nnn
{\cal O}[T_+]_{\sigma_+} U^{(J)}_m(\sigma)&=&{\lfloor
\varpi+J+m+1\rfloor \over \lfloor \varpi+1\rfloor}
U^{(J)}_m(\sigma)e^{ih(J+m)}
   {\cal O}[T_+]_{\sigma_+}\nnn
&-& \mu^{-1}{1\over \lfloor \varpi+1\rfloor}U^{(J)}_{m+1}(\sigma)
e^{-ih(\varpi-J-m-{3\over 2})}
{\lfloor J-m\rfloor \over \lfloor \varpi\rfloor \lfloor
\varpi-1\rfloor
}
{\cal O}[T_-]_{\sigma_+}.\nnn
&&
\label{Utr2}
\eeqa
Expanding to the first order in $J$ one finds\footnote{The
notation $\vartheta S^n(\sigma)$ means the regularized product of
of $\vartheta$ with $S^n$  at the point $\sigma$.}
\beqa
{\cal O}[T_-]_{\sigma_+} \vartheta S^n(\sigma)&=& \lfloor
\varpi\rfloor
\vartheta S^n(\sigma) e^{-ihn}{1 \over \lfloor \varpi\rfloor}
  {\lfloor \varpi-n-1\rfloor \over \lfloor \varpi-1\rfloor} {\cal
O}[T_-]_{\sigma_+}\nnn
&+& \mu \lfloor \varpi\rfloor \vartheta  S^{n-1}(\sigma)
e^{ih(\varpi-n-\demi)} \lfloor n\rfloor {\cal O}[T_+]_{\sigma_+},
\nnn
{\cal O}[T_+]_{\sigma_+} \vartheta S^n(\sigma)&=&  {\lfloor
\varpi+n+1\rfloor \over
\lfloor \varpi+1\rfloor}
\vartheta S^n(\sigma)e^{ihn}
   {\cal O}[T_+]_{\sigma_+}\nnn
&+&\mu^{-1}{1\over \lfloor \varpi+1\rfloor}  \vartheta
S^{n+1}(\sigma) e^{-ih(\varpi-n-{3\over 2})}
{\lfloor n\rfloor \over \lfloor \varpi\rfloor \lfloor \varpi-1\rfloor
}
{\cal O}[T_-]_{\sigma_+}\nnn
&+&\mu^{-1} \pi\alpha_-{\cos (hn)\over \sin h}S^{n+1}(\sigma)
e^{-ih(\varpi-n-{3\over 2})}
{1\over \lfloor \varpi\rfloor \lfloor \varpi-1\rfloor
}
{\cal O}[T_-]_{\sigma_+}.\nnn
&&
\label{thetatr}
\eeqa
In particular, for $n=0$ one obtains the transformation laws of the
B\"acklund field
\beqa
{\cal O}[T_-]_{\sigma_+} \vartheta(\sigma) &=& \lfloor \varpi\rfloor
\vartheta(\sigma) {1 \over \lfloor \varpi\rfloor}
  {\cal O}[T_-]_{\sigma_+}\nnn
{\cal O}[T_+]_{\sigma_+} \vartheta(\sigma) &=&
\vartheta(\sigma)
   {\cal O}[T_+]_{\sigma_+}
+\mu^{-1} \pi\alpha_-{1\over \sin h}S(\sigma) e^{-ih(\varpi-{3\over
2})}
{1\over \lfloor \varpi\rfloor \lfloor \varpi-1\rfloor
}
{\cal O}[T_-]_{\sigma_+}.\nnn
&&
\label{thetatr2}
\eeqa


\section{Operator  realization of  $U_q(sl(2)) \otimes U_q(sl(2))$}

In this section, we depart from Liouville theory and consider a
conformal theory
where the extended symmetry we just unravelled would be operatorially
realized.  The primary fields will be  assumed to be of the form
 $\Psi^{(J)}_{\rho \varpi}(\sigma)$. They are of the same type as the
Liouville
ones, but act on a larger Hilbert space. Their transformation laws
are
defined  by the usual co-product action, using the co-product
displayed on Eqs.\ref{coprodAB}, \ref{coprodom}. We are looking
for a linear map ${\cal O}$ from the matrices
$[A^\lambda],[B^\lambda],
[q^{\Omega_L}],[q^{\Omega_R}]$ to operators on the extended
Hilbert space, which generates this co-product action. Again,
${\cal O}$ will not be demanded to be an algebra homomorphism.
We thus require  for ${\cal O} ({\cal A}_{2\lambda})$ and
${\cal O} ({\cal B}_{2\lambda})$,
$$
{\cal O} ({\cal A}_{2\lambda})\Psi_{\rho ,\varpi} =
\sum _{\rho' ,\varpi'} \Psi_{\rho' ,\varpi'}
 \left[ [A^{-\lambda}]
_{\rho', \varpi';\,  \rho, \varpi} {\cal O}
({\cal A}_{2\lambda}) +
[B^{\lambda}]
_{\rho', \varpi';\,  \rho, \varpi}
 {\cal O}
({\cal B}_{2\lambda}) \right]
$$
\beq
{\cal O} ({\cal B}_{2\lambda})\Psi_{\rho ,\varpi} =
\sum _{\rho' ,\varpi'} \Psi_{\rho' ,\varpi'}
 \left[ [A^{\lambda}]
_{\rho', \varpi';\,  \rho, \varpi} {\cal O}
({\cal B}_{2\lambda}) +
[B^{-\lambda}]
_{\rho', \varpi';\,  \rho, \varpi}
 {\cal O}
({\cal A}_{2\lambda}) \right].
\label{defABO}
\eeq
Likewise, we would like to construct an operatorial realization
of the matrices $\Omega_L,\Omega_R$. The form of the co-product
Eq.\ref{coprodom} dictates that
$$
{\cal O} (q^{\Omega_L}) \Psi_{\rho ,\varpi} =
\Psi_{\rho ,\varpi} q^\rho {\cal O}(\hbox{\large 1})
$$
\beq
{\cal O} (q^{\Omega_R}) \Psi_{\rho ,\varpi} =
\Psi_{\rho ,\varpi} {\cal O} (q^{\Omega_R}).
\label{defOmeg}
\eeq
To simplify the formulae we dropped the superscript $(J)$
and the $\sigma$ dependences.
The latter are  similar to what we encountered
previously, and will be re-established
at the end.
Concerning $\Omega_R$, the above co-product structure is rather
particular in that the matrix $\Omega_R$ does not appear at all
in the linear actions on the $\Psi$ fields.
Let us note from the start that this co-action only implies that
${\cal O} (q^{\Omega_R}) $ commutes with all
the $\Psi$'s and  thus at this level only tells us that
${\cal O} (q^{\Omega_R}) $ is a central charge. To establish
a link with the matrix $\Omega_R$, we will impose later on
that the (FP) commutation relations of ${\cal O} (q^{\Omega_R}) $
with the other generators should reproduce the corresponding matrix
algebra.
 Concerning
operator products, we will assume that the
  matching condition Eq.\ref{mcdt}
holds, and only consider products of the type
$\Psi_{\rho_1 ,\varpi_1}\Psi_{\varpi_1 ,\varpi_2}$. Then it is easy
to verify that, on such products,   the
action just defined  has the same form as above, with the
matrices replaced by their co-products Eqs.\ref{coprodAB}
\ref{coprodom}.
This makes use of the special tensor product which obeys
Eq.\ref{match}, and
of the consistency relation
\beq
\Psi_{\rho ,\varpi} q^\varpi {\cal O}(\hbox{\large  1})=
\Psi_{\rho ,\varpi} {\cal O} (q^{\Omega_L}).
\label{coacmatch}
\eeq
In the above equation, contrary to the co-product actions to the
right,
 the matrix $\Omega_R$ does appear.
It is easy to check  that we may  consistently
assume that the
$\Psi$ fields satisfy the same fusion and
braiding relations  as the Liouville
$\psi$ fields.
Indeed, Eqs.\ref{fusinv} and \ref{newYB}
which were consequences  of the
covariance of the $\psi$ field operator algebra
 under ${\cal O} (T_\pm)$, imply the covariance
of the fusion and braiding of the $\Psi$ fields under the action of
${\cal O} ({\cal A}_{2\lambda})$, and ${\cal O} ({\cal
B}_{2\lambda})$.

The next step is to study what is the algebra
satisfied by the operators ${\cal O} ({\cal A}_{2\lambda})$,
${\cal O} ({\cal B}_{2\lambda})$, ${\cal O}(q^{\Omega_L})$ and ${\cal
O}
(q^{\Omega_R})$.
 Of course,   we are dealing with
FP relations similar to Eqs.\ref{alphaop} \ref{FPcom}, although this
is
hidden since
we do not write the $\sigma$ variables.
As expected we will find a suitable extension of the matrix
algebra Eqs.\ref{CAA}--\ref{relCAB}. First
using these
matrix relations, we derive from Eqs.\ref{defABO}, \ref{defOmeg},
\ref{coacmatch}
 the following
operatorial relations
\beq
{\cal O} (q^{\Omega _L} )\>{\cal O} ({\cal A}_{2\lambda})=
q^{2\lambda}
{\cal O} ({\cal A}_{2\lambda})\>  {\cal O} (q^{\Omega _L}),
\label{commu1}
\eeq
\beq
{\cal O} (q^{\Omega _L})\> {\cal O} ({\cal B}_{2\lambda})=
q^{-2\lambda}
{\cal O}({\cal B}_{2\lambda}) \> {\cal O} (q^{\Omega _L}),
\label{commu2}
\eeq
\beq
\left[ {\cal O} ({\cal A}_{2\lambda}) ,{\cal O} ({\cal
B}_{2\lambda}) \right]
= C^{\lambda}_1 {\cal O} (\lfloor \Omega _L \rfloor ),
\label{commu3}
\eeq
\beq
 {\cal O} ({\cal A}_{2\lambda})\> {\cal O} ({\cal A}_{-
2\lambda})
- {\cal O} ({\cal B}_{2\lambda})\> {\cal O} ({\cal B}_{
-2\lambda})
= C^{\lambda}_2 {\cal O} (\lfloor \Omega _L \rfloor ).
\label{commu4}
\eeq
where
the $C^{\lambda}_i$ are central terms which commute
with the $\Psi$ fields. Let us note that
any operator that commutes with the $\Psi$'s automatically
commutes with ${\cal O}(q^{\Omega_L})$ as well, as a
trivial consequence of the first of Eqs.\ref{defOmeg}.
Thus
\beq
[C^{\lambda}_i, {\cal O}(q^{\Omega_L})]=0
\label{omLcomm}
\eeq
 We do not assume, however, that the
$C_i^\lambda$ commute with the other generators, and it will turn
out that in fact they don't.  Moreover we find that the
quadratic operator $\left[ {\cal O} ({\cal A}_{2\lambda}) ,
{\cal O} ({\cal
 B}_{-2\lambda}) \right]$
and another operator noted $X^{\lambda} ({\cal A},{\cal B})$
transform among  themselves under the co-product action.
The operator  $X^{\lambda} $  is given  by
$$
X^{\lambda} ({\cal A},{\cal B}) =
\left( {\cal O} (\lfloor \Omega _L -2\lambda \rfloor
\right)^{-1} Z^{\lambda} ({\cal A},{\cal B}) -
\left( {\cal O} (\lfloor \Omega _L +2\lambda \rfloor \right)^
{-1} Z^{-\lambda} ({\cal A},{\cal B}),
$$
where
$$
Z^{\lambda} ({\cal A},{\cal B}) =
{\cal O} ({\cal A}_{2\lambda})\> {\cal O} ({\cal A}_{-2
\lambda})
- {\cal O} ({\cal B}_{-2\lambda})\> {\cal O} ({\cal B}_
{2\lambda}).
$$
In order to proceed further, we have to make ans\"atze.
First, it is easy to see from the definitions  that
${\cal O} ({
\cal A}_{2\lambda})$
and ${\cal O} ({\cal B}_{-2\lambda})$ act the same way,
 the  matrices involved
 being the same. It is therefore natural to postulate  that
\beq
\left[ {\cal O} ({\cal A}_{2\lambda}) ,{\cal O} ({\cal
B}_{-2\lambda}) \right] = 0.
\label{commu5}
\eeq
Second, by analogy with the matrix algebra, we assume that
\beq
{\cal O}(q^{\Omega_R}) {\cal O} ({\cal A}_{2\lambda})=
{\cal O} ({\cal A}_{2\lambda}) F^\lambda \!
\left ( {\cal O}(q^{\Omega_R})\right).
\label{hyp1}
\eeq
\beq
{\cal O}(q^{\Omega_R}) {\cal O} ({\cal B}_{2\lambda})=
{\cal O} ({\cal B}_{2\lambda})
G^\lambda \!  \left ( {\cal O}(q^{\Omega_R}\right ),
\label{hyp2}
\eeq
where $F$ and $G$ are unknown  functions of one variable.
The co-product action on $\Psi$ (Eqs.\ref{defABO}, \ref{defOmeg})
 implies that $F^\lambda=G^\lambda$. Note that
the operator ${\cal O}(q^{\Omega_R})$ may still be replaced by
an arbitrary function, say $K$,  of itself since we only know that it
commutes with all the $\Psi$'s. This replacement
is equivalent to changing
\beq
 F^\lambda\left ( {\cal O}(q^{\Omega_R})\right) \to
K\left( F^\lambda\left ( K^{-1}\left (
{\cal O}(q^{\Omega_R})\right)\right)\right),
\label{tr}
\eeq
where $ K^{-1}$ is the inverse function.
We will only consider only the class of functions $F$ for which
 there exists  a function $K$ such that Eq.\ref{tr}
gives $F^+(x)\to xq$. This choice is precisely what is needed
to reproduce the matrix algebra on the operatorial level (up to
central terms).
 Passing $ {\cal O} (q^{\Omega _R})$
from left to right on both sides of Eq.\ref{commu4}, and making use
of
Eqs.\ref{hyp1} \ref{hyp2}, one sees that $F^\lambda$ obey the
relation
$F^\lambda\left(F^{-\lambda}(x)\right)=x$.
Thus there exists  a redefinition of
$ {\cal O} (q^{\Omega _R})$ such that
$F^\lambda(x)\to xq^{2\lambda}$. We adopt this choice hereafter,
so that   Eqs.\ref{hyp1}, \ref{hyp2} become
\beq
{\cal O}(q^{\Omega_R})\> {\cal O} ({\cal A}_{2\lambda})=
q^{2\lambda}\> {\cal O} ({\cal A}_{2\lambda})
 \>  {\cal O}(q^{\Omega_R}),
\label{commu8}
\eeq
\beq
{\cal O}(q^{\Omega_R})\>  {\cal O} ({\cal B}_{2\lambda})=
 q^{2\lambda} \> {\cal O} ({\cal B}_{2\lambda})
   \>  {\cal O}(q^{\Omega_R}).
\label{commu9}
\eeq

Let us now prove  that
 the relations just written
 allow us to derive
the   operator algebra, and that
 it coincides  with
the one we obtained for matrices (Eqs.\ref{CAA} -- \ref{relCAB})
up to central terms. First
Eq.\ref{commu5}  implies that
$ X^{\lambda} ({\cal A},{\cal B}) = 0$ and
making use of Eq.\ref{commu4} one finds
\beqa
\left[ {\cal O} ({\cal A}_{2\lambda}) ,{\cal O} ({\cal
A}_{-2\lambda}) \right] &=&
{1\over q +q^{-1}} \left( C^{\lambda}_2\> {\cal O} (\lfloor \Omega _L
-2\lambda \rfloor ) -  C^{-\lambda}_2\> {\cal O} (\lfloor \Omega _L
+2\lambda \rfloor ) \right)\nnn
&&\label{commu6}\\
\left[ {\cal O} ({\cal B}_{2\lambda}) ,{\cal O} ({\cal
B}_{-2\lambda}) \right] &=&
{1\over q +q^{-1}} \left( C^{-\lambda}_2\> {\cal O} (\lfloor \Omega
_L
-2\lambda \rfloor ) - C^{\lambda}_2\> {\cal O} (\lfloor
 \Omega _L
+2\lambda \rfloor ) \right)\nnn
&& \label{commu7}
\eeqa
In general,  we must verify that
our assumptions are   consistent with higher commutators (typically
 Jacobi identities).
By the same argument as for Eq.\ref{omLcomm},  we have
\beq
\left [{\cal O}(q^{\Omega_R}),\, {\cal O}(q^{\Omega_L})\right ]
=0.
\label{hyp3}
\eeq
Next,
passing $ {\cal O} (q^{\Omega _R})$
from left to right on both sides of Eq.\ref{commu3}, and making use
of
Eqs.\ref{hyp3} \ref{commu8} \ref{commu9}
one sees that
$$
C_1^\lambda=0.
$$
Furthermore,
commuting ${\cal O}(q^{\Omega_R})$ with both sides of
Eq.\ref{commu4},
and taking Eqs.\ref{hyp3} and \ref{commu8}f into account, one obtains
\beq
\left [C_2^\lambda ,\, {\cal O}(q^{\Omega_R})\right ]
=0
\label{hyp4}
\eeq
Now let us consider the Jacobi identity between
${\cal O} ({\cal A}_{2\lambda})$, ${\cal O} ({\cal A}_{-2\lambda})$,
${\cal O} ({\cal B}_{2\lambda})$. Since we now know that
the last operator commutes with the first and with the second, this
gives
$$
\left[ \left[{\cal O} ({\cal A}_{2\lambda}),\,
{\cal O} ({\cal A}_{-2\lambda})\right],
{\cal O} ({\cal B}_{2\lambda})\right]=0,
$$
and, according to Eq.\ref{commu4},
$$
\left[ \left( C^{\lambda}_2\> {\cal O} (\lfloor
\Omega _L
-2\lambda \rfloor ) - C^{-\lambda}_2\> {\cal O} (\lfloor
 \Omega _L
+2\lambda \rfloor ) \right),\> {\cal O} ({\cal B}_{2\lambda})
\right ]=0.
$$
Exchanging the role of $A$ and $B$, one also deduces that
$$
\left[ \left( C^{-\lambda}_2\> {\cal O} (\lfloor
\Omega _L
-2\lambda \rfloor ) - C^{\lambda}_2\> {\cal O} (\lfloor
 \Omega _L
+2\lambda \rfloor ) \right),\> {\cal O} ({\cal A}_{2\lambda})
\right ]=0.
$$
In order to solve these equations, we look for operators that
commute with ${\cal O} ({\cal B}_{2\lambda})$ or
$ {\cal O} ({\cal A}_{2\lambda})$. Combining Eqs.\ref{commu8},
\ref{commu9} with Eqs.\ref{commu1}, and \ref{commu2}, respectively,
one deduces that
$$
\left[{\cal O}(q^{\Omega_R-\Omega_L}),\,
{\cal O} ({\cal A}_{2\lambda})
\right]  =0,\qquad
\left[{\cal O}(q^{\Omega_R+\Omega_L}),\,
{\cal O} ({\cal B}_{2\lambda})
\right]  =0
$$
Thus we are lead to make the ansatz
$$
{1 \over q+q^{-1}}
\left( C^{+}_2\> {\cal O} (\lfloor
\Omega _L
-1 \rfloor ) - C^{-}_2\> {\cal O} (\lfloor
 \Omega _L
+1 \rfloor ) \right)=
c^+{\cal O}(q^{\Omega_R+\Omega_L})-
c^-{\cal O}(q^{-\Omega_R-\Omega_L})
$$
$$
{1 \over q+q^{-1}}
\left( C^{-}_2\> {\cal O} (\lfloor
\Omega _L
-1 \rfloor ) - C^{+}_2\> {\cal O} (\lfloor
 \Omega _L
+1 \rfloor ) \right)=
d^+{\cal O}(q^{\Omega_R-\Omega_L})-
d^-{\cal O}(q^{-\Omega_R+\Omega_L})
$$
where $c^\pm$, and $d^\pm$ commute with all operators, and
$C_2^\lambda$
is taken to be independent of ${\cal O}(q^{\Omega_L})$. Note that
these last relations could not involve higher powers of
${\cal O}(q^{\pm(\Omega_R+\Omega_L)})$, since by assumption the
left hand sides are  linear in ${\cal O}(q^{\pm\Omega_L})$.
Solving the two above equations, one finds that $d^\pm=c^\pm$, and
\beqa
C_2^+&=&c^+q{\cal O} (q^{\Omega _R})-
c^-q^{-1}{\cal O} (q^{-\Omega _R}),\nnn
C_2^-&=&c^+q^{-1}{\cal O} (q^{\Omega _R})-
 c^-q{\cal O} (q^{-\Omega _R}).
\label{central}
\eeqa
Finally, substituting these last two relations in Eqs.\ref{commu4},
one finally derives the operator algebra.
It is given by
$$
\Bigl [ {\cal O} ({\cal A}_{2\lambda}),
\, {\cal O} ({\cal B}_{2\mu}) \Bigr]=
\Bigl [ {\cal O}(q^{2 {\cal A}_3}), {\cal O}(q^{2 {\cal B}_3})
 \Bigr]=0
$$
$$
\Bigl [ {\cal O} ({\cal A}_{2\lambda}),
\,  {\cal O}(q^{2 {\cal B}_3}) \Bigr]=
\Bigl [  {\cal O}(q^{2 {\cal A}_3}), {\cal O} ({\cal B}_{2\mu})
 \Bigr]=0,
$$
$$
{\cal O}(q^{2 {\cal A}_3}) {\cal O}({\cal A}_{2\lambda}) =
q^{4\lambda} {\cal O}({\cal A}_{2\lambda}) {\cal O}(q^{2 {\cal
A}_3}),
$$
$$
\Bigl [ {\cal O} ({\cal A}_{+}),
\, {\cal O} ({\cal A}_{-}) \Bigr]= c^+ {\cal O}(q^{2 {\cal A}_3}) -
c^- {\cal O}(q^{-2 {\cal A}_3}),
$$
$$
{\cal O}(q^{2 {\cal B}_3}) {\cal O}({\cal B}_{2\lambda}) =
q^{4\lambda} {\cal O}({\cal B}_{2\lambda})
 {\cal O}(q^{2 {\cal B}_3}),
$$
$$
\Bigl [ {\cal O} ({\cal B}_{+}),
\, {\cal O} ({\cal B}_{-}) \Bigr]= c^+ {\cal O}(q^{2 {\cal B}_3})  -
c^- {\cal O}(q^{-2 {\cal B}_3}),
$$
$$
{\cal O} ({\cal A}_{+}){\cal O} ({\cal A}_{-})
-{c^+q {\cal O}(q^{2 {\cal A}_3})+
c^-q^{-1} {\cal O}(q^{-2 {\cal A}_3})\over q-q^{-1}}=
$$
\beq
{\cal O} ({\cal B}_{+}){\cal O} ({\cal B}_{-})
-{c^+q {\cal O}(q^{2 {\cal B}_3})+
c^-q^{-1} {\cal O}(q^{-2 {\cal B}_3})\over q-q^{-1}}.
\label{comfin}
\eeq
We have defined, as for matrices,
$$
{\cal O}(q^{2 {\cal A}_3})= {\cal O}(q^{\Omega_R}) {\cal O}
(q^{\Omega_L}),
$$
\beq
{\cal O}(q^{2 {\cal B}_3})= {\cal O}(q^{\Omega_R}) {\cal O}
(q^{-\Omega_L}).
\label{O3def}
\eeq

The next  step  is
to discuss the $\sigma$ dependence. In analogy
with the case of Liouville (for $U_{\sqrt{q}}(sl(2))$), it is natural
to
assume that $ {\cal O} ({\cal A}_{\pm}) $ $ {\cal O} ({\cal B}_{\pm})
$
depend upon one point (have gradation one),
and that $ {\cal O} (q^{\Omega _L})$,
 $ {\cal O} (q^{\Omega _R})$ have gradation zero.
Then the central terms $c^\pm$
have gradation two, that is, would be written explicitly
 as $c^\pm(\sigma_1,\sigma_2)$. With this, it is straightforward
to re-establish the $\sigma$ dependences. We will not do it
explicitly. The commutators written above are actually not true
ones, but instead are similar to the left hand side of
Eq.\ref{FPcom}.
One may verify that the FP version of Jacobi identity holds, so that
our discussion indeed makes sense.

In order to make a closer contact with our Liouville discussion, let
us show finally
that we may construct generators of the $U_{\sqrt{q}}(sl(2))$
algebra also in the present framework. Taking $c^+=c^-=c$ for
simplicity
one defines
\beq
{\cal O}(T_\pm)=
\left ( {\cal O}(\lfloor \Omega_R/2\rfloor )\right)^{-1}
\left ({\cal O} ({\cal A}_{\pm})+{\cal O} ({\cal B}_{\mp})\right).
\label{Tgendef}
\eeq
Making use of the commutation relations derived above one deduces
\beqa
\Bigl [ {\cal O} (T_{+}),
\, {\cal O} (T_{-}) \Bigr]&=&c'\>
 {\cal O} (\lfloor \Omega_L\rfloor),  \nnn
{\cal O}(q^{\Omega_L})\> {\cal O} (T_{\pm})&=&
q^{\pm 1} {\cal O} (T_{\pm})\> {\cal O}(q^{\Omega_L}).
\label{Tgencom}
\eeqa
It is easily seen that the co-product action of the operators just
defined is the same as the action of the  $U_{\sqrt{q}}(sl(2))$
generators in Liouville theory (Eq.\ref{2TactAB}). This is a
consistent
co-product action since   it follows from the co-product definition
Eq.\ref{coprodAB} that, for matrices
\beq
\Lambda (T_{2\lambda}) = A^{\lambda} \underbrace \otimes
T_{2\lambda}  +
B^{-\lambda}  \underbrace \otimes  T_{-2\lambda}.
\label{coactT}
\eeq
\section{Novel Hopf like  algebraic structure}
As  we already mentioned,  our internal symmetry algebra
 $U_q(sl(2))\otimes U_q(sl(2))$ does not
obey the usual axioms of a Hopf algebra, since in  particular it does
not admit a counit in the usual sense. However,
in this last section we show that it does
possess  an algebraic structure which is a natural  generalization
of the
Hopf algebra  axioms.   For orientation  let us recall them. Let
${\cal G}$ be a
Hopf algebra. It is equipped with   a multiplication $m$,  a
comultiplication $\Lambda$,
an  antipode $s : {\cal G}\to {\cal G}$ and a counit
$\epsilon : {\cal G}\to {\cal C}$ ( ${\cal C}$  the
set of complex numbers), with the following properties
\beq
m(a\otimes 1)=m(1\otimes a)=a,\quad
m(m\otimes {\rm id})=m( {\rm id}\otimes m)=m,
\label{mprop}
\eeq
\beq
(\Lambda\otimes {\rm id})\Lambda=
({\rm id}\otimes\Lambda)\Lambda,
\quad
\Lambda(a)\Lambda(b)=\Lambda(ab)
\label{deltaprop}
\eeq
\beq
(\epsilon\otimes {\rm id})\Lambda=
({\rm id}\otimes \epsilon) \Lambda ={\rm id}
\label{epscoprod}
\eeq
\beq
\epsilon(ab)=\epsilon(a)\epsilon(b)
\label{epsmult}
\eeq
\beq
s(ab)=s(b)s(a)
\label{smult}
\eeq
\beq
\Lambda(s)=(s\otimes s)P\Lambda
\label{scoprd}
\eeq
\beq
m(s\otimes {\rm id})\Lambda(a)=
m({\rm id}\otimes s)\Lambda(a)=\epsilon(a).1
\label{msprop}
\eeq
where $P$ is the permutation operator.
Let us recall that our algebra may be expressed in two equivalent
ways, that is, either in terms of
the generators $A_\pm$, $B_\pm$ $q^{\varpi_R}$, $q^{\varpi_L}$, or in
terms of the
generators ${\cal A}_\pm$ ${\cal B}_\pm$, $q^{{\cal A}_3}$, $q^{{\cal
B}_3}$. Each set has its
virtues and drawbacks, so that they should be used according to the
question
addressed. A general generator  will be
denoted $K^a$ or ${\cal K}^a$ depending upon the description chosen.
For the coproduct, which was
defined in section 4, we have a more standard form in terms of the
$K$ generators. The
corresponding structure
constant $\Lambda ^a_{bc}$ will be defined such that
\beq
\Lambda({ K}^a) =\Lambda ^a_{bc} { K}^b {\underbrace \otimes} { K}^c.
\label{stctL}
\eeq
Due to the matching condition, this does not uniquely specify
$\Lambda ^a_{bc}$. However,
some of the equations we will check later on remove this ambiguity.
The appropriate choice is
that Eq.\ref{coprodAB} and \ref{coprodom} hold directly. However, for
the unit element of
${\cal G}$, we have to
define the structure constant such that
\beq
\Lambda({\bf 1})=q^{-\varpi_R}{\underbrace \otimes}q^{\varpi_L}.
\label{coprod1}
\eeq
This is consistent since the right hand side is equal to
$ {\bf 1}{\underbrace \otimes}{\bf 1}$, but $\Lambda ^a_{bc}$ should
be defined
without making this replacement\footnote{
Note that this is also consistent with the coproduct action due to
Eq.\ref{coacmatch}.}
. Next Eqs.\ref{deltaprop} mean that the coproduct is
coassociative and preserves the algebra. These properties were
verified in section 4
using the ${\cal K}$ form, where however one cannot define structure
constants
similar to $\Lambda ^a_{bc} $.
Let us next discuss the counit. From Eq.\ref{smult}, and the fact
that $\epsilon$   is a number,  we see that it  defines a
one dimensional representation of ${\cal G}$.
Eq.\ref{epscoprod} means that
  that its    coproduct  with any other representation
gives back the same  representation only---hence for the usual
$U_q(sl)2))$, the
counit is simply the one dimensional spin zero representation.
However, as we already observed, in our case the relevant spin zero
representation is infinite dimensional.
{}From the viewpoint of conformal theory, this  is natural, since the
corresponding
$\psi_{\rho, \omega}^{(0)}$ is proportional to the identity operator
in the
Hilbert space, while a
one dimensional representation for ${\bf{\cal A}}$ would correspond
to a
projector onto
a single Verma module. Since the fusion of the identity operator with
any other
$\psi$ gives
back the same operator, one
sees that the present infinite dimensional representation with $J=0$
should play
 the role of
counit, and we next show that this is true.
Since the identity operator does not shift the zero modes, it
is consistent
that this counit of a novel type
  be restricted to  $\rho=\varpi$. Thus, making use of Eq.\ref{ABexp}
we define the counit as given by
the spin zero representation, with ${ B}=0$.
This  may be written compactly, for a general element ${K}^a$,  as
\beq
\epsilon ( K^{a})_{\rho,\varpi,\rho',\varpi'}
= \epsilon ( K^{a})_{\varpi} \delta ( K^{a})_{\varpi,\varpi'}
 \delta_{\rho,\varpi} \delta_{\rho',\varpi'}
\label{epsilonred}
\eeq
where
\beq
\epsilon ( { A}_{\pm    })_{\varpi} = \pm 1, \quad
\epsilon ( { B}_{\pm    })_{\varpi} = 0, \quad
\epsilon ( q^{\varpi_R} )_{\varpi} = q^{\varpi},\quad
\epsilon ( q^{\varpi_L} )_{\varpi} = q^{\varpi},\quad
\label{couexp}
\eeq
\beq
\delta ({\cal A}_{\pm    } )_{\varpi,\varpi'}= \delta_{\varpi',\,
\varpi\pm 1}=
\delta ({\cal B}_{\pm    } )_{\varpi,\varpi'},\quad  \delta (q^{{\cal
A}_3})_{\varpi,\varpi'}=
\delta ( q^{{\cal B}_3} )_{\varpi,\varpi'}= \delta_{\varpi',\,
\varpi}.
\label{deltdef}
\eeq
One may  verify that the coproduct of this representation with any
spin $J$
representation
gives a single representation with the same spin $J$. Indeed on has
(the upper indices specify the representations)
\beq
\Lambda\left
({\bf{\cal K}^a}\right)^{(J_1,0)}
_{\rho_1', \varpi_1', \rho_2',\varpi_2';\,  \rho_1, \varpi_1,
\rho_2,\varpi_2 }
=\left[{\bf{\cal K}^a}\right]^{(J_1)}
_{\rho_1'\, \varpi_1';\, \rho_1,\, \varpi_1}
\delta_{\rho'_2,\varpi'_2}\delta(K^a)_{\varpi_2,\varpi'_2}
\delta_{\rho_2,\varpi_2}\delta_{\rho_2,\varpi_1},
\label{counr}
\eeq
with similar equations after reversing the orders.
It is easy to see that the right hand side of the equation just
written
satisfies the
same spin $J_1$ relations as the matrices we started from, although
it acts
 non
trivially
in the second space. Note that a trivial action in the second space
would have
been inconsistent
since our ${\underbrace \otimes}$ definition which appears in the
coproduct
respects
the matching condition Eq.\ref{mcdt}, so that, if indices are shifted
in the
first space,
there must be also some shift in the second. The equations just
written are the analoga of
Eqs.\ref{epscoprod}. It is easy to see that the left and right hand
sides contain the same factors,
so that we may rewrite in general, making use of the form
Eq.\ref{epsilonred},
\beq
\Lambda ^a_{bc} ({ K}^b)_{\rho,\varpi,\rho',\varpi'}
\epsilon ( { K}^{c})_{\varpi '}=
\Lambda ^a_{bc} \epsilon ( { K}^{b})_{\rho '}
 ({ K}^c)_{\rho,\varpi,\rho',\varpi'}
= ({ K}^a)_{\rho,\varpi,\rho',\varpi'},
\label{epscoprodL}
\eeq
which is very similar to Eq.\ref{epscoprod}.
The only difference is that, here,
 $\epsilon$ has
an index.

Let us now turn to the antipode. As discussed in our first paper
along the same line\cite{CGS1},
it appears when one goes from the right-action which we
have been using so far here
 to the left-action. This is neatly done
by making use of Eqs.\ref{defABO}, but
one may  also define the antipode by using
the right-action of ${\cal O}(T_\pm)$,
since Eq.\ref{coactT} shows that it acts by the same coproduct
coefficient and matrices $A^\lambda$ and $B^\lambda$ as
${\cal O}({\cal A}_{2\lambda})$
(or ${\cal O}({\cal B}_{-2\lambda})$). Moreover,
the left-action can be derived either directly by inverting relations
  Eq.\ref{TactAB}, or Eq.\ref{defABO},
 or else by hermitian conjugation of Eq.\ref{TactAB}.
In any case,
one arrives at the following
definition of the antipode, which it is simpler to handle  in terms
of ${\cal K}^a_{(S)}$
operators
\beq
{\cal A}_{\pm \,(S)} = -(\lfloor\Omega_R\rfloor)^{-1}
{\cal A}_{\mp}\lfloor \Omega_R\rfloor,  \quad
{\cal B}_{\pm \,(S)}  = -(\lfloor\Omega_R\rfloor)^{-1}
{\cal B}_{\pm }\lfloor \Omega_R\rfloor
\label{ABSdef}
\eeq
\beq
q^{{\cal A}_3}_{ (S)} = q^{{\cal A}_3},\quad
q^{{\cal B}_3}_{ (S)} = q^{-{\cal B}_3}.
\label{omSdef}
\eeq
 Next,  making use of Eqs.\ref{coprodAB}, \ref{coprodom} and the
matching condition, one may verify that
$$
\Lambda (A^{\lambda}_{(S)}) =
A^{\lambda}_{(S)} \underbrace \otimes A^{\lambda}_{(S)} +
B^{\lambda}_{(S)}  \underbrace \otimes B^{-\lambda}_{(S)}
$$
$$
\Lambda (B^{\lambda}_{(S)}) =
B^{\lambda}_{(S)} \underbrace \otimes A^{-\lambda}_{(S)} +
A^{\lambda}_{(S)}  \underbrace \otimes B^{\lambda}_{(S)}
$$
\beq
\Lambda(q^{\Omega_L}_{(S)})={\bf 1}\underbrace\otimes
q^{\Omega_L}_{(S)} \qquad
 \Lambda(q^{\Omega_R }_{(S)})= q^{\Omega_R }_{(S)} \underbrace
\otimes {\bf 1}.
\label{scoprodL}
\eeq
These are the analoga
 of Eq.\ref{scoprd}, since they take the same form as the
coproduct of the $K$ generators, with the two factors exchanged.
 On the other hand, the antipode just defined take the
general form ${\cal K}^a_{(S)}=U^{-1} S^a_b {\cal K}^b U$ where
$S^a_b$ are  very simple
constants. This
immediately allows us  to verify the analoga
 of Eq.\ref{smult}, namely that ${\cal K}^a_{(S)}$
satisfies the same algebra as ${\cal K}^a$ if one reverses the order.
 Finally, one may check
that
\beqa
\Lambda ^a_{bc} ( K^c K^{b}_{(S)})_{\rho,\varpi,\rho',\varpi'}
 &=& \epsilon ( K^{a})_{\varpi } \delta _{\rho \rho '} \delta_{\varpi
\varpi '}, \nnn
\Lambda ^a_{bc} ( K^{c}_{(S)} K^b )_{\rho,\varpi,\rho',\varpi'}
 &=& \epsilon ( K^{a})_{\rho } \delta _{\rho \rho '} \delta_{\varpi
\varpi '},
\label{mspropL}
\eeqa
which are the analoga of Eq.\ref{msprop}.
\section{Conclusion}
The idea of using the primary fields  in  the smallest quantum group
representation as generators has led to interesting developments.
In our previous article, we recovered in this way the
well-known quantum group symmetries of the covariant operator
algebra;
however, nontrivial central terms were  found to appear in the
corresponding
generator algebra, and the free field zero mode was seen to play an
unusual
role. In this paper, we have analyzed the more familiar Bloch
wave/Coulomb gas operator basis of conformal field theory using the
same idea.
The nonlinear character of the transformation relating the two bases
has been seen to lead to a surprisingly different realization of the
quantum
group symmetry for the Bloch waves, with a novel $U_q(sl(2))\otimes
U_q(sl(2))$ structure with somewhat unusual properties emerging. It
represents the underlying symmetry of the operator algebra exactly in
the same way as $U_q(sl(2))$ did for the covariant basis. Our
approach is  constructive in the sense that we
 deduce the symmetry structure directly
from the operator algebra under consideration, without a priori
assumptions.
We were thus lead to modify some of the standard Hopf algebra axioms
in order to accommodate the properties of the above structure.
In particular, the  coproduct
prescription differs from the standard one by the presence of  two
additional constraints (matching condition and equality of
Casimir eigenvalues), which incorporate rather naturally the CFT
features
of the theory. Moreover, we were led to define
the counit in terms of an infinite dimensional spin zero
representation
rather than a map to the complex numbers,   since it corresponds to
the identity operator in the operator algebra. In this way
we arrived at a self-consistent symmetry structure, the
representation theory of which reproduces the spectrum of operators
of the theory.

Of course there remain many open questions, either of a mathematical
or
physical nature. Let us list some of them:
\begin{itemize}
\item A more systematic mathematical understanding of our
$U_{q}(sl(2))\otimes U_q(sl(2))$
 and $U_{\sqrt{q}}(sl(2))$ structures
is clearly desirable.
\item We have discussed only briefly  the case where $q$ is a root
of unity, the case of rational conformal field theory, where
interesting subtleties are expected to appear.   A systematic
treatment
of the Coulomb gas picture in this case is provided by Felder's
formalism \cite{F}, which would have to be adapted to the slightly
different
formulation of vertex operators in the present framework.
\item
One would like to understand better the CFT framework of section 6
 for the operatorial realization of our extended symmetry
group.
\item Our discussion differs strongly from previous analyses of
quantum
group symmetries in rational CFT, and the connection is not obvious.
\item So far we have not considered the question of how our
generators
act on the
Hilbert space of states, rather than the covariant fields of the
theory.
 There is a double enigma here: First, the connection between states
and
operators is nontrivial in view of the  problem of the $SL(2)$ -
invariant
vacuum (in the CFT sense) \cite{LuS}. Second,  the existence
a vacuum state invariant under the quantum group as postulated
in \cite{MaSc},
assumes that the counit is of the usual type, i.e. a complex
number. This problem must be reexamined from scratch with our
counit of a novel type.
\item In general one would like to put our results to some practical
use
in solving Liouville theory. One obvious application would be to
use the quantum group generators for the classification of
observables
in the strong coupling theory \cite{G3}, where no classical
interpretation
is available, and quantum group invariance becomes the sole defining
property
of observables.
\item It would be interesting  to see how
our analysis extends to the higher Toda theories where the quantum
group symmetry has a higher rank.
\end{itemize}

\noindent{\bf Acknowledgements}
We would like to thank A. Alekseev for interesting discussions.
One of us (J.-L. G.) is grateful to CERN for its  financial
support, and warm hospitality.

\end{document}